# Programmable Photonic Unitary Processor Enables Parametrized Differentiable Long-Haul Spatial Division Multiplexed Transmission

Mitsumasa Nakajima[1,a,*], Kohki Shibahara[2,*], Kohei Ikeda[3,4,*], Akira Kawai[2], Masaya Notomi[3,4], Yutaka Miyamoto[2], and Toshikazu Hashimoto[1]

1. NTT Device Technology Laboratories, 3-1 Morinosato-Wakamiya, Atsugi, Kanagwa, Japan

2. NTT Network Innovation Laboratories, 1-1 Hikarinooka, Yokosuka, Kanagwa, Japan

3. NTT Basic Research Laboratories, 3-1 Morinosato-Wakamiya, Atsugi, Kanagwa, Japan

4. NTT Nanophotonics Center, 3-1 Morinosato-Wakamiya, Atsugi, Kanagwa, Japan

    a. *mitsumasa.nakajima@ntt.com*

*\* Equally contributed to this paper*

The explosive growth of global data traffic demands scalable and energy-efficient optical communication systems. Spatial division multiplexing (SDM) using multicore or multimode fibers is a promising solution to overcome the capacity limit of single-mode fibers. However, long-haul SDM transmission faces significant challenges due to modal dispersion, which imposes heavy computational loads on digital signal processing (DSP) for signal equalization. Here, we propose parameterized SDM transmission, where programmable photonic unitary processors are installed at intermediate nodes. Instead of relying on conventional digital equalization only on the receiver side, our approach enables direct optimization of the SDM transmission channel itself by the programmable unitary processor, which reduces digital post-processing loads. We introduce a gradient-based optimization algorithm using a differentiable SDM transmission model to determine the optimal unitary transformation. As a key enabler, we first implemented telecom-grade programmable photonic unitary processor, achieving a low-loss (2.1 dB fiber-to-fiber), wideband (full C-band), polarization-independent, and high-fidelity ($R^2$>96% across the C-band) operation. We experimentally demonstrate 1300-km transmission using a three-mode fiber, achieving strong agreement between simulation and experiment. The optimized photonic processor significantly reduces modal dispersion and post-processing complexity. Our results establish a scalable framework for integrating photonic computation into the optical layer, enabling more efficient, high-capacity optical networks.

The explosive growth of global network traffic, fueled by high-bandwidth applications, such as ultra-high-definition video streaming, cloud computing, and mobile communications, has driven the continuous advancement of optical backbone networks. While wavelength-division multiplexing (WDM) combined with digital coherent detection has extended the capacity of single-mode fiber (SMF), the transmission system is now approaching the nonlinear Shannon limit [1–3]. Spatial division multiplexing (SDM) with multicore or multimode fiber (MMF) offers a path forward by enabling parallel spatial channels to dramatically enhance spectral efficiency without increasing the physical footprint [1,2,4–6]. MMF provides the highest spatial information density, and long-haul mode-division multiplexing (MDM) transmission has already demonstrated impressive capacity under standard cladding constraints [7–11]. However, SDM systems face severe impairments such as inter-mode crosstalk and differential mode delay (DMD), which result in complex signal distortions that cannot be separated spectrally. These distortions are typically compensated via multiple-input multiple-output digital signal processing (MIMO-DSP), but the complexity scales as $O(N^2L)$, with $N$ modes and time dispersion window $L$ for compensating DMD. This scaling rapidly becomes impractical for long-haul transmission, where $L$ reaches hundreds of nanoseconds [4]. While several prior studies have explored fixed mode-scrambling [12,13] or mode-permutation [11] techniques to suppress modal dispersion, these methods are limited in flexibility and adaptability. Most rely on passive, static reconfigurations and often require heuristic or trial-and-error optimization to identify effective configurations. As the number of spatial modes increases, such combinatorial strategies become impractical, and they cannot accommodate dynamic changes in fiber conditions.

A fundamental limitation of conventional SDM implementations is that the optical-domain transmission matrix is treated as fixed and uncontrollable. This leads to an inherent inefficiency: MIMO-DSP must compensate for fully distorted signals at the receiver, requiring enormous computational resources, while no intermediate optimization is possible along the transmission path. In contrast, the wireless communications field has developed hybrid analog-digital architectures to mitigate the complexity of massive MIMO systems by incorporating analog preprocessing units such as a Butler matrix and metasurface to reduce DSP loads [14−16]. More recently, photonic processors have been explored for RF-domain signal processing, which offer significant improvements in power efficiency and footprint reduction [17-20] by leveraging the advances of photonic computing platforms for machine-learning and quantum computations [21−28]. Similarly, photonic processing for a very short-reach MMF transmission (~1 meter) [29] and free-space optical link (~10 cm) has been demonstrated [30]. These trends suggest that integrating programmable photonic processing into SDM fiber networks could provide means to reduce DSP complexity by pre-processed signals at the optical domain.

Despite the promise of optical-domain MIMO-assisted processing, there have been no prior demonstrations of photonic-assisted MIMO equalization for long-haul fiber transmission, primarily due to the following major technical challenges. Assuming photonic processors are applied for MIMO equalization, the required optical hardware would scale as $O(N^2L)$, leading to an impractically large number of interferometric elements, especially for long-haul transmission where $L$ is large, e.g., typically, $L>100$ is required for the equalization of ten-nanosecond order dispersion. Current photonic processors, which typically consist of tens to hundreds of Mach-Zehnder interferometers (MZIs) [19,22−26], lack the scalability required for such implementations. In addition, SDM fiber networks operate across multiple multiplexing dimensions, including space, wavelength, and polarization over transmission distances of hundreds to thousands of kilometers, requiring photonic processors to preserve fidelity across a broad transmission bandwidth with low insertion loss. Typical photonic processors have largely been designed for single-wavelength operation, with only limited experimental demonstrations across broad optical spectrum, making them unsuitable for practical deployment in SDM/WDM-compatible networks.

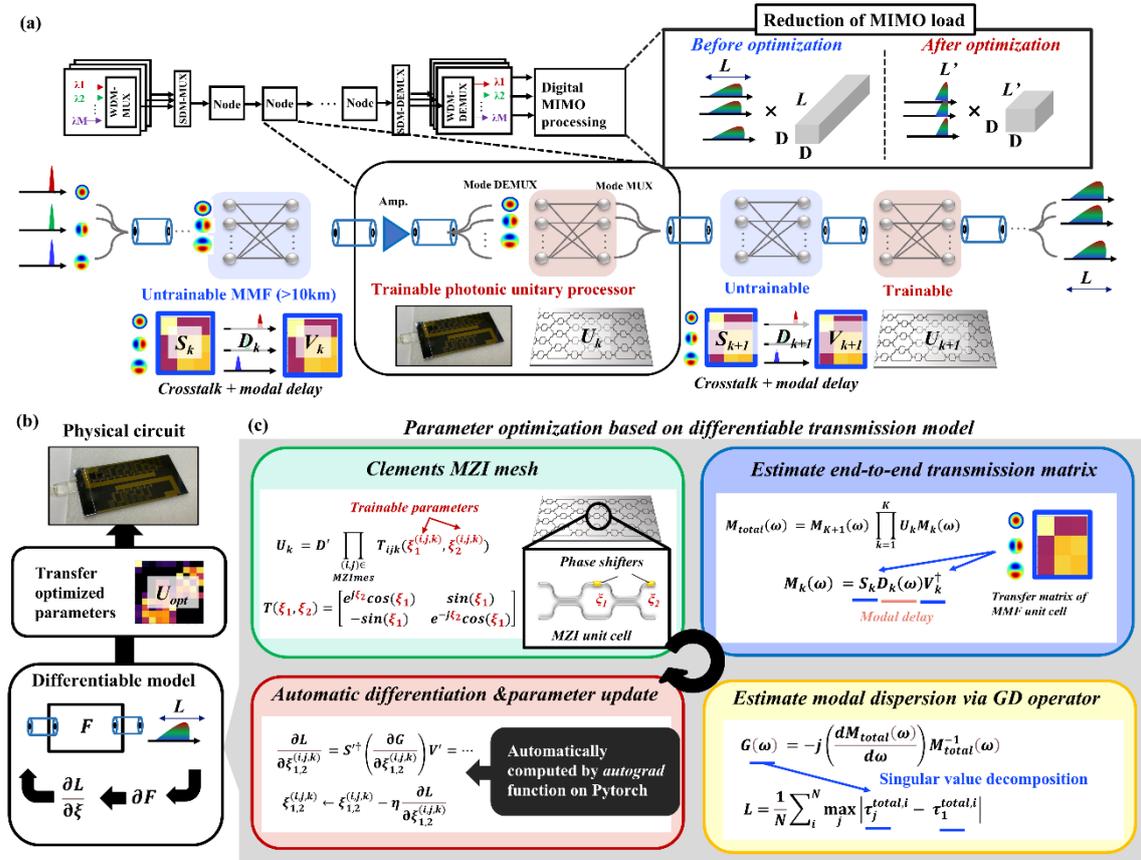

*Figure 1. Parameterized differentiable SDM transmission with photonic unitary processors.* (a) Schematic explanation of parametrized SDM transmission. Statistical characteristics can be optimized by using trainable photonic unitary processors enabling direct optimization of the transmission matrix to reduce the pre/post MIMO processing load. (b) Optimization of photonic unitary conversion based on differential propagation model. (c) Concrete step of optimization: trainable phase parameters $\xi$ in the photonic unitary processors are optimized by estimating the gradient $\partial L/\partial \xi$.

Here, we present the first demonstration of long-haul *parameterized SDM transmission*, in which programmable MZI-based photonic unitary processors are installed at intermediate fiber nodes to manipulate the transmission matrix in-line. It enables the compensation of accumulated signal distortions at each stage, thereby significantly improving scalability compared to traditional receiver-side-only compensation approaches. To find an optimum photonic unitary transformation, we introduce a gradient-based optimization algorithm based on a differentiable model of the SDM transmission system. Leveraging machine learning techniques, we confirm the feasibility and scalability of the approach across various multimode fiber conditions. Using our proposed algorithm and high-quality photonic processor, we experimentally realize a parametrized SDM transmission over 1300 km in a three-mode fiber system. As a key enabler, we implement a telecom grade photonic processor based on a silica-based planar lightwave circuit (PLC) platform [31, 32], achieving low-loss (2.1 dB fiber-to-fiber), broadband (full C-band), polarization-independent, and high-fidelity (>96% across the C-band) operation. Our results show excellent agreement between simulation and experiment, validating the practicality and effectiveness of our approach. Our approach establishes a scalable framework for integrating photonic computation into the optical layer, enabling more efficient, high-capacity optical networks.

**Results**

**Parameterized SDM-transmission with differentiable physical model**

Figure 1(a) shows a conceptual diagram of the proposed system, named parameterized SDM transmission, where programmable photonic unitary processors are installed within each SDM transmission node. Unlike conventional optical SDM transmission, where modal dispersion is mitigated solely by digital MIMO processing at the receiver, our approach directly optimizes the optical channel itself by configuring the installed photonic unitary circuit. This enables end-to-end optimization of physical transmission characteristics, surpassing conventional analog MIMO methods at the transmitter or receiver side. Our primary optimization target is the differential mode delay of the transmission link, as it directly impacts DSP complexity in a linear manner. Importantly, DMD is a relatively static characteristic of the fiber transmission system, unlike fast-varying impairments such as phase noise or polarization fluctuations. Thus, the compensation of DMD does not require high-speed tracking, allowing the optimized unitary transformations to remain effective over long timescales. This property makes our approach highly practical for real-world deployment, as the photonic circuits can operate with minimal control overhead. Figure 1(b) illustrates an overview of the optimization processing. In the proposed scheme, we introduce a differentiable transmission model of the SDM transmission link (denoted as $F$ in the figure), allowing gradient-based optimization of photonic unitary processor parameters to minimize modal dispersion by calculating the differential $\partial F$. After the optimization within the digital twin, we transfer the optimal unitary matrix ($U_{opt,k}$) to the physical circuit at the $k$th transmission span ($k = 1, 2, 3, …, K$).

Figure 1(c) shows the considered transmission model of the parameterized SDM transmission link. In the model, we first define the photonic $N×N$ unitary matrix $U_k$ supporting with $N$ spatial modes using trainable phase parameters $\xi_1^{(i,j,k)}$ and $\xi_2^{(i,j,k)}$ within MZIs. The unitary conversion $U_k$ can be calculated as follows: $U_k = D' \prod_{(i,j) \in MZI-mesh} T_{ijk}(\xi_1^{(i,j,k)}, \xi_2^{(i,j,k)})$, where $i$ and $j$ denote the position of MZI in the mesh, $D'$ is a diagonal matrix, and $T_{ijk}$ is the 2×2 unitary conversion [33]. The transmission matrix of the MMF with $D$ spatial modes can be defined as $M_k(\omega) = S_k D_k(\omega) V_k^\dagger$, where $S_k$ and $V_k$ represent the input and output unitary mode coupling and $D_k(\omega)$ denotes the diagonal matrix, meaning the modal delay within each fiber span [34,35]. The accumulated modal delay $\tau^{total}$ corresponds to the singular value of the group delay operator, $G(\omega)$, by using the end-to-end frequency-dependent transfer matrix $M_{total}(\omega)$ [35]. The maximum DMD can be defined as $max_j|\tau_j^{total} - \tau_1^{total}|$ ($j$=1, 2, …, $N$), which directly affects the temporal window length $L$ of MIMO-DSP processing (the detailed mathematical descriptions are given in the Method section). Notably, the entire process to compute the modal delay is based on linear algebra, and it is fully differentiable. Thus, we leverage this property for backpropagation-based gradient optimization, allowing for effective reduction of modal dispersion by systematically tuning the unitary transformations. The detailed steps of the optimization is also described in the Method section. In the following section, we set the same unitary matrix for every span, meaning $U_k$=$U$, as a first proof of concept.

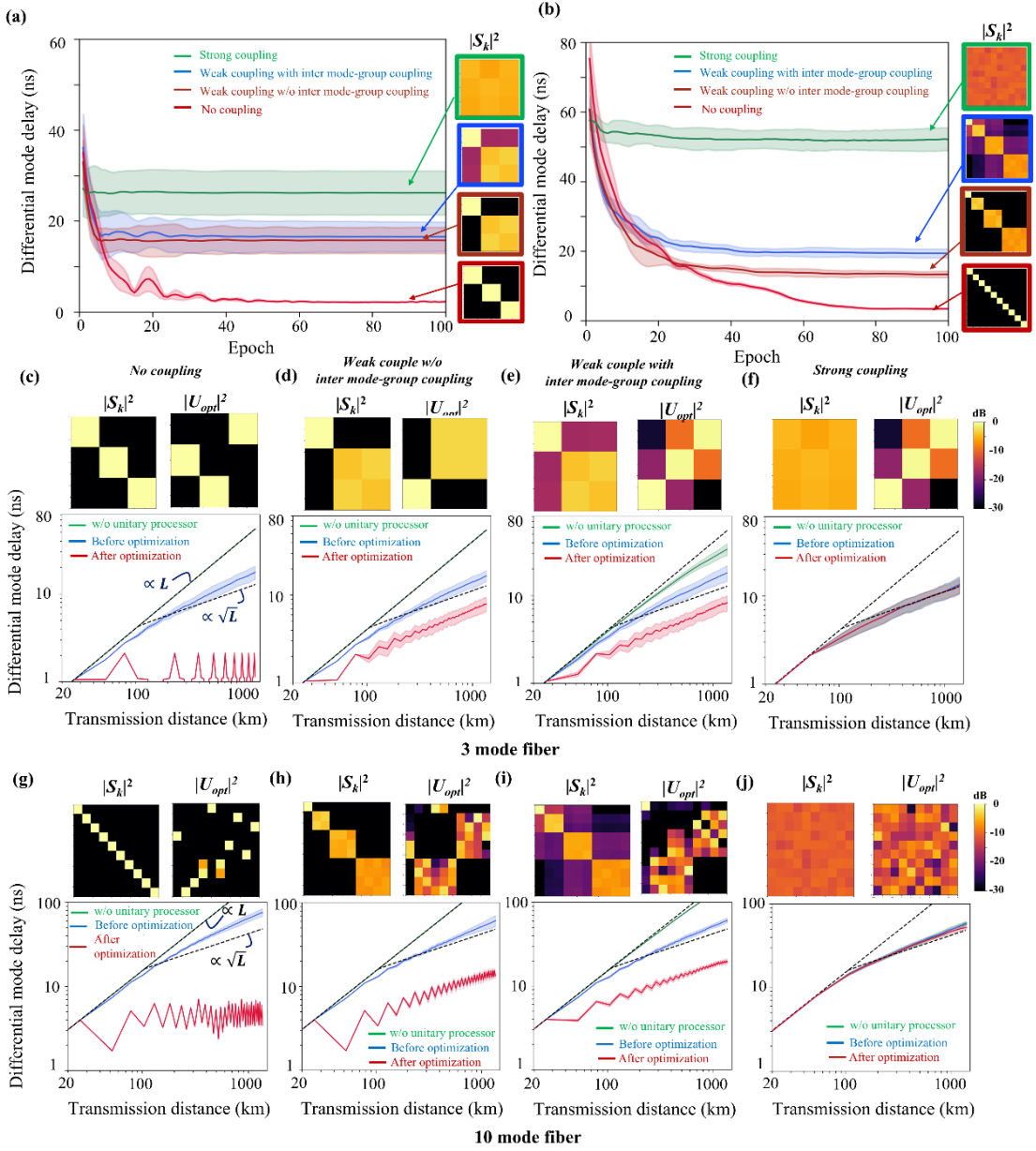

*Figure 2. Optimization of unitary matrix to reduce modal dispersion of MMF.* Convergence process of modal dispersion over training epochs for (a) three and (b) ten- mode fibers under different coupling conditions of the MMF. Modal dispersion as a function of transmission distance for (c)–(f) three and (g)–(j) ten- mode fibers before (i.e., random matrixes drawn as blue lines) and after (red line) optimization of unitary matrices with four possible MMF coupling scenario; (c), (g) no coupling, (d), (h) strong coupling within degenerated modes, (e), (i) strong coupling within degenerated modes and weak coupling with inter mode-group, and (f), (j) strong coupling within all modes. The upper part of each figure shows defined MMF transmission matrixes and optimized unitary matrices. All filled areas denote standard deviations based on 100 simulation trials with changing random seeds for generation phase perturbation and MMF matrix.

To assess the impact of our optimization approach, we executed numerical simulations for the parametrized SDM transmission over graded-index MMFs supporting three and ten spatial modes. To investigate different mode coupling scenarios, four distinct conditions were evaluated: (i) no coupling, (ii) weak coupling within mode groups only, (iii) weak coupling both within and across mode groups, and (iv) strong coupling across all modes. Figure 2(a) and (b) shows the training curves of parameterized SDM transmission in three-mode and ten-mode MMF. As shown in Fig. 2(a) and (b), modal dispersion converges as training epochs increase, confirming that the optimized unitary transformation matrix significantly reduces the DMD. The inset images show the coupling matrices under different conditions, revealing distinct convergence characteristics depending on the strength of intermodal coupling. Figure 2(c)–(j) show the modal dispersion before and after optimizing the unitary matrix $U$ for three-mode and ten-mode fibers. As references, we also plot the case without any unitary processors. In the worst-case scenario, the DMD accumulates linearly with distance (dashed line noted as $\propto L$ in the figures), corresponding to the case where both $U_k$ and $M_k(\omega)$ are the identity matrices. When $M_k(\omega)$ or $U_k$ are the random unitary matrices [see Fig. 2(f,j) and the blue line in Fig. 2(c)–(j)], a statistical averaging suppresses the evolution of the modal dispersion like a random walk process, leading to the modal dispersion with scaling as the square root of distance [12,13,34]. As shown in Figs. 2(f,j), all the plots agree with the $\propto \sqrt{L}$ lines, suggesting that the effective unitary conversion does not exist for the MMF with strong coupling. On the other hand, for the no/weak coupling MMF cases, we could find the unitary matrices outperforming the DMD for the random matrices for DMD. This suggests that unitary matrix optimization can suppress DMD accumulation rather than only having a simple ensemble averaging effect. The optimization effect was found to be most significant in the absence of the modal coupling, and it gradually approached the performance of random coupling as the strength of coupling increased. We also found that denser deployment of intermediate photonic unitary processors along the transmission path enhances the effectiveness of modal dispersion compensation (see Supplementary Information S1).

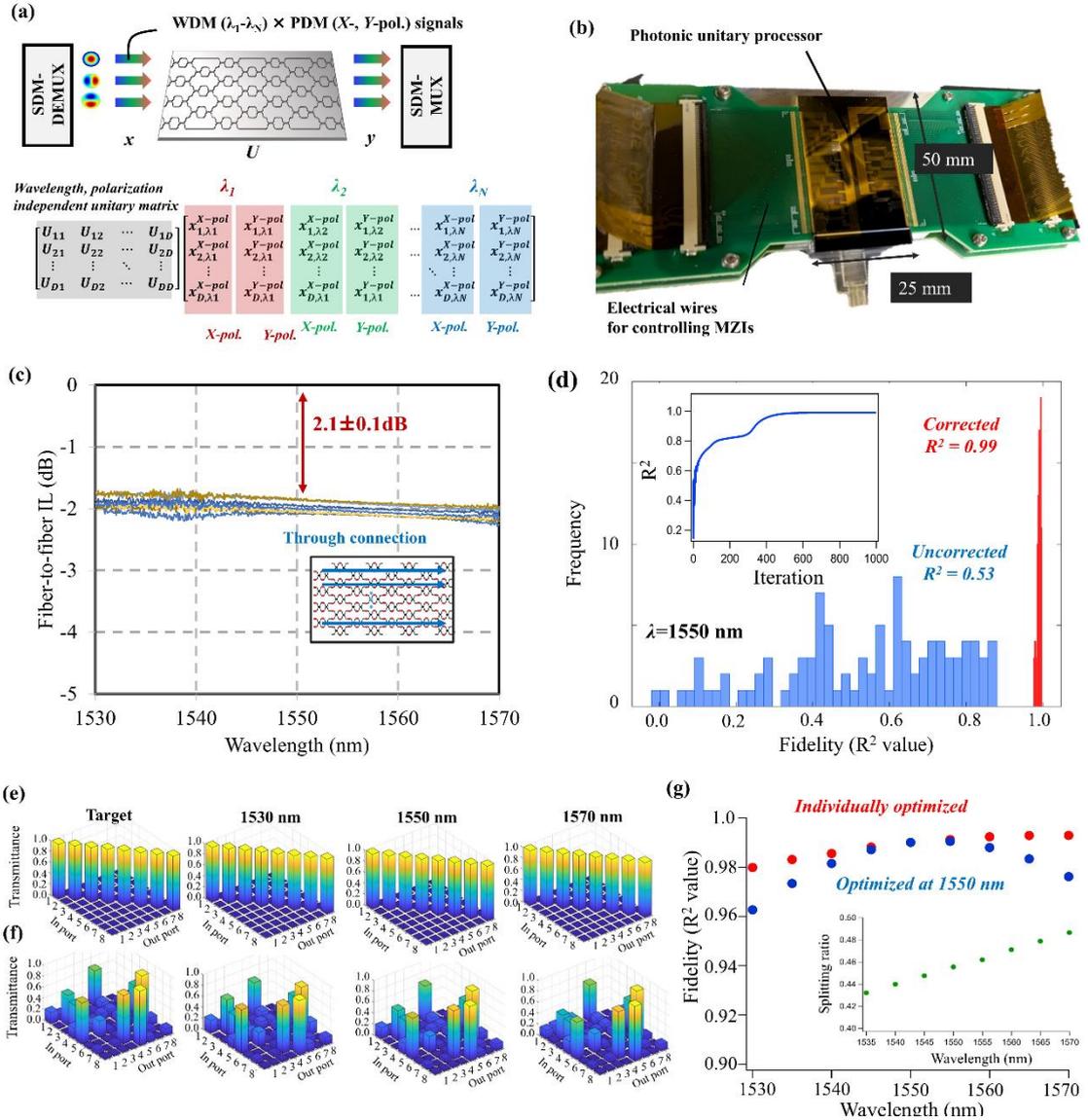

*Figure 3. PLC-based photonic unitary processor for SDM-WDM-PDM transmission.* (a) Schematic of photonic unitary processors required for parametrized SDM transmission. The photonic processor should execute unitary operation insensitive to the wavelength and polarization state, as the standard SDM transmission utilizes not only space division but also wavelength and polarization division. (b) Photograph of fabricated PLC-based Clements-type photonic unitary circuit (50 mm × 25 mm). (c) Fiber-to-fiber insertion loss as a function of wavelength. Color difference corresponds to difference of input/output ports.. (d) Fidelity histogram at 1550 nm before and after calibration (inset: learning convergence curve). e), (f) Target and measured transmission matrix at 1530, 1550, and 1570 nm for (e) identity and (f) random unitary matrix. (g) Wavelength dependence of fidelity, comparing individually optimized (red) and optimized at 1550 nm (blue). The inset shows the estimated splitting ratio of the directional coupler in the MZI.

**SDM-WDM-PDM compatible programable photonic unitary processor**

Figure 3(a) shows a schematic explanation of the photonic unitary circuit required for parametrized SDM transmission. As the SDM signals typically support both WDM and polarization division multiplexed (PDM) signals, a high signal-to-noise-ratio (SNR), low-loss, broadband, polarization-independent, and high-fidelity photonic unitary processor is required for the demonstration of the parametrized SDM transmission. Here, we focused on a silica-based PLC technology [31,32]. PLCs have already been installed in optical fiber links, quantum photonic chips, and an optical lattice clock thanks to their wide passband (visible to infrared), high stability (over ten years), and fabrication tolerance [37–40]. Therefore, the PLC platform can better meet the requirements for long-haul SDM transmission than conventional silicon-based implementations can, which typically suffer from insertion loss and wavelength and polarization dependence. The comparison with a photonic processor on a silicon platform is summarized in Supplementary Information S2.

Here, we fabricated an 8×8 Clements-type MZI mesh based on PLC technology for parametrized SDM transmission [see Fig. 3(b)]. The chip size is 25 mm × 50 mm [41].(details of the fabrication and the basic characteristics of the PLC-based MZI-mesh are described in the Method section). By carefully designing the Clements configuration [33] without light path differences, we mitigated wavelength dependence caused by delay-path interferometry while reducing port-dependent loss. To compensate for the residual errors remaining after deterministic calibration, we performed machine learning-based calibration, which infers the unknown parameters using the dataset obtained during the implementation of random matrices. The detailed calibration step is described in Supplementary Information S3. Figure 3(c) shows the fiber-to-fiber insertion loss of the fabricated photonic unitary processor as a function of wavelength. Each color reflects the difference of the input/output ports. As can be seen, we could achieve fiber-to-fiber loss of 2.1 dB with almost complete insensitivity to wavelength and spatial ports. This loss is far superior to that for previous photonic unitary processors; typically, >10 dB fiber-to-fiber loss and single-wavelength operation [21,22,29,30]. Also, our circuit does not require polarization diversity, unlike silicon photonic circuits [42].

Figure 3(d) shows a histogram of fidelity values at 1550 nm before and after machine-learning based calibration. The uncorrected PLC, meaning that we utilized the control parameter without any initial phase errors in each MZI arm, showed inferior fidelity, exhibiting an $R^2$ value of 0.53, where $R^2 = 1 - \sum_i |y_i^{measured} - y_i^{ideal}|^2 / |y_i^{ideal}|^2$. After the proposed calibration, $R^2$ improves to 0.99, indicating that the compensation scheme effectively mitigates systematic errors. The inset of figure shows the training curve under the calibration. The monotonic increase of $R^2$ suggests the successful operation of the proposed calibration method. Figure 3(e) and (f) show examples of measured transmittance matrix $|U|^2$ at 1530, 1550, and 1570 nm after the calibration, compared with the target matrix for the identity (upper) and random (lower) matrix. All the observed matrices showed the same shapes as the target matrices, suggesting successful operation for wideband operation. Figure 3(g) shows the wavelength dependence of the fidelity, evaluated by comparing the experimentally obtained outputs at each wavelength with the model output constructed and optimized at 1550 nm (blue dots). Although the fidelity gradually degraded from the center wavelength, $R^2$ > 0.96 was maintained in the whole telecom C-band. From the noise dependency analysis shown in Supplementary Information S4, we confirmed $R^2$ > 0.95 is one criterion to suppress the performance degradation in the parameterized SDM transmission. Thus, we consider the observed fidelity to be sufficient for the purpose of this study. To confirm the reason for the remaining wavelength dependence, we individually inferred the splitting ratio of the directional coupler for each wavelength using our calibration method [see the red dots in the Fig. 3(h); the inset of figure shows inferred wavelength dependency of the splitting ratio]. Afterwards, the fidelity improved to more than 0.98 over the C-band, suggesting that the wavelength dependence of couplers is the main factor in the observed degradation. Thus, we can further improve the fidelity by utilizing wavelength-insensitive splitters such as multimode-interferometers (MMIs).

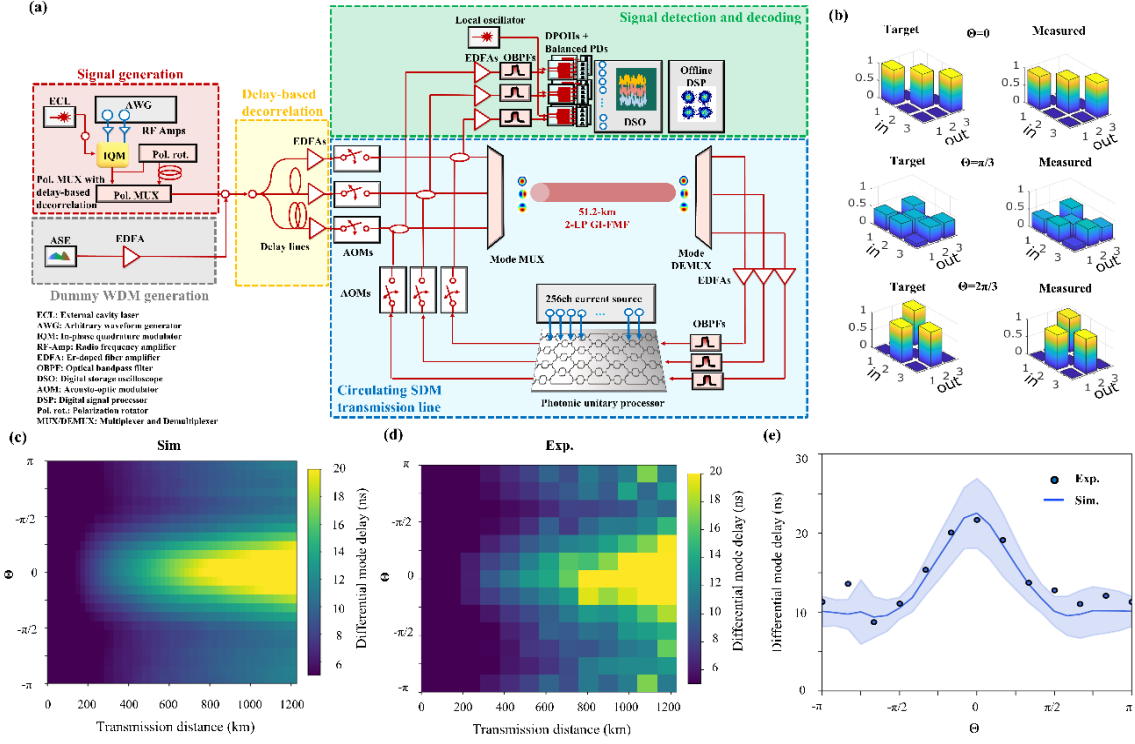

*Figure 4. Experimental setup and basic characteristics of parameterized SDM transmission.* (a) Schematic diagram of the experimental setup for long-haul three-mode fiber transmission, incorporating a photonic unitary processor. (b) Comparison of simulated and experimental mode dispersion as a function of Θ. (e) Implemented unitary matrices for different Θ setups. θ. (c) Numerical simulation and (d) experimental results of mode delay as a function of transmission distance and unitary transformation parameter Θ. (e) Mode delay accumulation over transmission distance for different Θ (Θ=0, -π, -2π/3), comparing experimental (dot plots) and simulation (slid lines) results. In the transmission experiment, we sent 33,360 symbols and measured the impulse response through digital MIMO postprocessing and defined its 95% region as the differential mode delay. The simulations were repeated 100 times, and averaged values are plotted. The shaded region in (e) corresponds to the standard deviation of the simulation.

**Experimental Demonstration of Long-Haul Parametrized SDM Transmission**

To demonstrate the proof-of-concept of the parametrized SDM transmission, we constructed an experimental setup [Fig. 4(a)]. In this setup, modulated 12-GBaud PDM quadrature-phase-shift-keying (QPSK) signals were circulated in the graded-index MMF-based fiber loop via acoustic optical modulators (AOMs), which enables path switching for the arbitrary timing of signal circulation. The MMF had 51.2-km length, supporting three-mode transmission. The setup enables the investigation of the transmission characteristics with varying transmission spans (i.e., number of recirculations within the MMF loop). Our experimental setup corresponds to testing the span-independent unitary matrix $U_k = U$ for the three-mode fiber transmission shown in Fig. 1. The experimental total modal d ispersion $L$ was quantitatively analyzed by the required MIMO equalizer window. A more detailed explanation is described in the Method section.

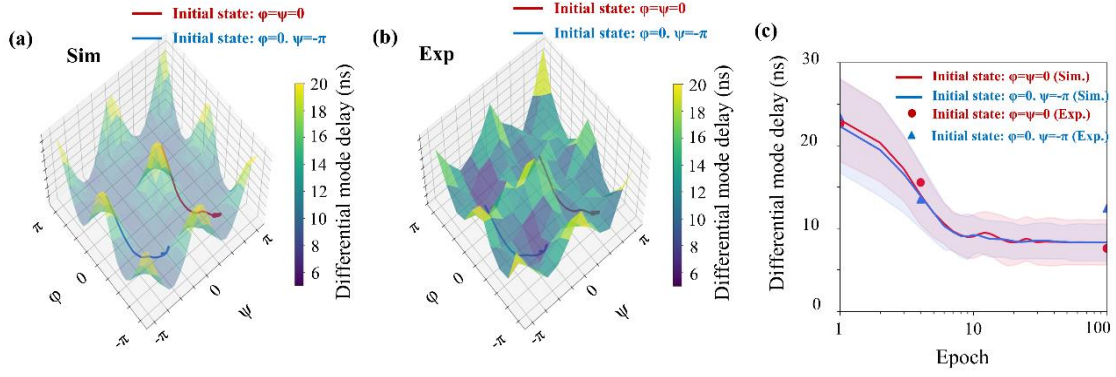

*Figure 5. Optimization using differentiable SDM transmission model. (a), (b) Potential landscape of (a) simulated and (b) experimental mode delay as a function of trainable parameters. Solid lines show the trajectory under the optimization step landscape for different initial states. (c) Training curve of modal dispersion for simulation (solid lines) and experiment (circle and triangle plot). The simulations were repeated 100 times, and averaged values are plotted. In the transmission experiment, we sent 33,360 symbols and measured the impulse response through digital MIMO postprocessing and defined its 95% region as the differential mode delay. The shaded region in (c) corresponds to the standard deviation of the simulation.*

First, to investigate the validity of our transmission model shown in Fig. 1(c), we compared the numerical and experimental results using deterministically designed 3 × 3 unitary matrices $U_{det}(\Theta)$, which are characterized by a rotation angle $\Theta \in [-\pi, \pi)$ around the axis of $\left(\frac{1}{\sqrt{3}}, \frac{1}{\sqrt{3}}, \frac{1}{\sqrt{3}}\right)$ as follows [43].

$$U_{det}(\Theta) = \frac{1}{3}\begin{pmatrix} 1 + 2cos(\Theta) & 1 + 2cos(\Theta + 2\pi/3) & 1 + 2cos(\Theta - 2\pi/3) \\ 1 + 2cos(\Theta - 2\pi/3) & 1 + 2cos(\Theta) & 1 + 2cos(\Theta + 2\pi/3) \\ 1 + 2cos(\Theta + 2\pi/3) & 1 + 2cos(\Theta - 2\pi/3) & 1 + 2cos(\Theta) \end{pmatrix}. \quad (1)$$

With $\Theta = 0$, $U_{det}(\Theta)$ works as the identity matrix. The set of $\Theta = \pm 2\pi/3$ produces a permutation matrix, which enables continuous change from the identity to permutation matrix via a random state, effectively [see Fig. 4(b)]. Figure 4(c) and (d) show the differential mode delay as a function of transmission distance and the $\Theta$. Both results exhibit a clear dependence on $\Theta$, confirming that the in-line optical unitary processor could control the DMD. While the experimental data in Fig. 4(d) exhibit non-smooth fluctuations, which could be caused by experimental non-idealities such as phase fluctuation in the transmission line, the key trends align well between simulation and experiment. Figure 4(e) shows a quantitative comparison between simulated and experimental data for mode delay variation with $\Theta$. The shaded area indicates the standard deviation of the simulation results. The good agreement between the experiment and simulation results suggests the validity of our constructed digital twin model. The detailed dependency of the impulse response on the $\Theta$ value is summarized in Supplementary Information S5.

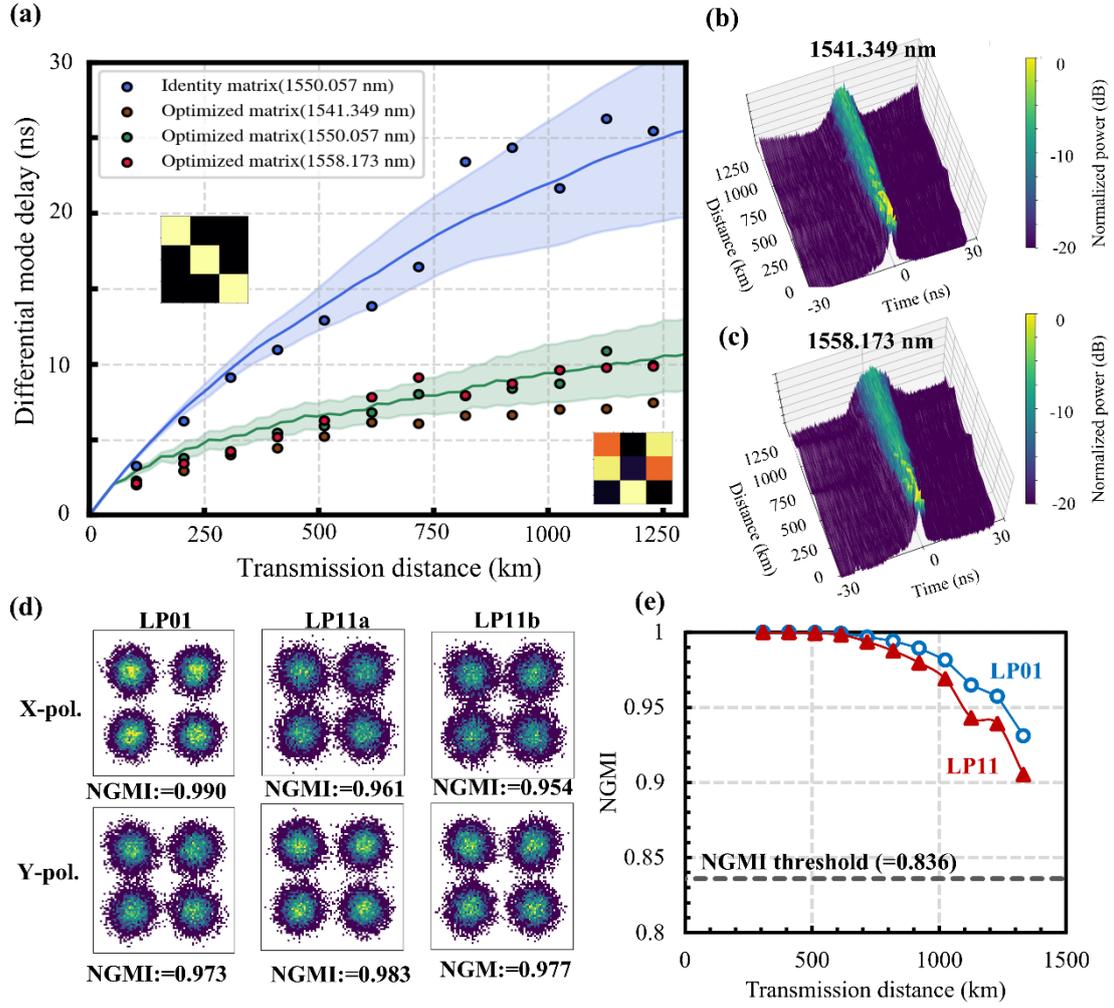

*Figure 6. Transmission characteristics of parameterized SDM transmission with optimized photonic unitary processing.* (a) Mode delay accumulation over transmission distance for different before (identity matrix) and after optimization, comparing experimental (dot plots) and simulation (slid lines) results. For the experimental results, the results for wavelengths of 1541.349-, 1550.057-, and 1558.173-nm are also plotted to confirm the adaptivity to the WDM signals. (b), (c) evolution of Impulse response along transmission distance for (b) 1541.349 and (c) 1558.173 nm. (d) Received constellation after 1024-km transmission for each spatial and polarization modes. The estimated NGMI values are shown in the lower part. (e) NGMI as a function of transmission distance. The dashed line indicates error-free threshold of 0.836. In the transmission experiment, we sent 33,360 symbols and measured the impulse response through digital MIMO postprocessing and defined its 95% region as the differential mode delay. The simulations were repeated 100 times, and averaged values are plotted. The shaded region in (c) corresponds to the standard deviation of the simulation.

Next, we investigated whether our constructed differential model could optimize the differential mode delay to reduce the digital MIMO load. Through the simulation, we observed that different optimal unitary matrices appeared depending on the initial conditions of each trial (see Supplementary Information S6). This suggests that multiple solutions exist due to the nature of the optimization landscape. To better visualize how the optimization parameters evolve within the unitary matrix, we introduced the following 3D rotation matrix with three independent rotation parameters, $\theta$, $\varphi$, and $\psi$, to simplify the optical circuit parameters:

$$U_{rot}(\theta, \varphi, \psi) =$$

$$\begin{pmatrix} \cos(\theta)\cos(\varphi) & -\sin(\theta)\cos(\varphi)+\cos(\theta)\sin(\varphi)\sin(\psi) & \sin(\theta)\sin(\varphi)+\cos(\theta)\sin(\varphi)\cos(\psi) \\ \sin(\theta)\cos(\varphi) & \cos(\theta)\cos(\varphi)+\sin(\theta)\sin(\varphi)\sin(\psi) & -\cos(\theta)\sin(\varphi)+\sin(\theta)\sin(\varphi)\cos(\psi) \\ -\sin(\varphi) & \cos(\varphi)\sin(\psi) & \sin(\varphi)\sin(\psi) \end{pmatrix}. \quad (2)$$

As described in Supplementary Information S7, we found that $\theta$ exhibits minimal variation, meaning that for this three-mode fiber system, parameters $\varphi$ and $\psi$ primarily determine the optimized unitary transformation. Figure 5(a) shows the potential landscape of the simulated differential mode delay as a function of $\theta$ and $\varphi$ values. The trajectories under the backpropagation-based optimization are also shown in the figure. The different trajectories correspond to different initial conditions. It was found that there are multiple optimized points minimizing the mode delay for the considered three-mode fiber case, which could be caused by degenerate feature of the MMF: i.e., $LP_{11a}$ and $LP_{11b}$ are in equilibrium. Also, the optimized parameters are highly dependent on the initial trainable parameters, indicating the optimizer searches for the nearest optimal parameters from the initial states. Therefore, the reason for the obtained initial parameter dependency could be this potential surface. Figure 5(b) presents the experimentally obtained potential landscape of the differential mode delay as a function of $\theta$ and $\varphi$ values. The trajectory shows simulated trajectory under the optimization. Compared with Figs. 5(a) and (b), the experimentally measured landscape closely resembles the digital twin prediction. Therefore, the trajectory could find the minimal points even for the experimental potential landscape. Figure 5(c) compares the simulated and experimentally obtained differential mode delay evolution under the training epoch. The experimental results well agree with the simulated prediction. These results validate the gradient-based optimization framework using the proposed differential transmission model.

Finally, we investigated whether the optimized unitary transformation derived from the digital twin model suppresses modal dispersion as predicted. Here, we executed the transmission experiment by implementing the optimized unitary matrix on our photonic processor. Figure 6(a) shows the experimental (dot plot) and simulated (solid line) accumulated differential mode delay as a function of transmission distance for wavelengths of 1541.349, 1550.057, and 1558.173 nm. For reference, the case for the identity matrix, which corresponds to the case without the any effective unitary transformation, is also plotted in the figure. While there were multiple optimal unitary matrices as discussed above, we implemented the optimized unitary matrix shown in the inset of Fig. 6(a) in this experiment. As can be seen in the figure, the results well agree with the expectation from the simulated one. Compared to the identity matrix (the case without unitary transformation), the optimized transformations exhibit significant suppression from ~25 to ~10 ns for all wavelengths, which highlights the successful wideband and high-fidelity operation of our fabricated photonic unitary processor. As the length of the differential mode delay linearly affects the digital MIMO-DSP load, the results indicates that we could reduce digital post-processing complexity compared with the case without photonic unitary processor. Figures 6(b) and 6(c) illustrate the evolution of the impulse response along the transmission distance for two representative wavelengths. The results confirm that the impulse responses were maintained over the entire transmission distance for each wavelength. Figure 6(d) presents constellation diagrams for each spatial and polarization mode after 1024-km transmission. The estimated normalized generalized mutual information (NGMI) [44] values for corresponding spatial and polarization modes are also shown in the figure. All constellations show well-clustered QPSK points and NGMI values well above the error-free threshold of 0.836, reflecting the low insertion loss and polarization insensitivity of our fabricated photonic unitary circuit. Figure 6(e) shows the NGMI as a function of transmission distance up to 1331.2 km. The NGMI values were maintained above 0.836 over the whole transmission distance over the 1300 km [45]. These experimental results highlight the potential of our proposed parametrized SDM transmission approach using the wideband, polarization-insensitive, and high-fidelity unitary processor enabled by our silica-based PLC Clements mesh.

**Discussion**

A key advantage of the proposed parameterized SDM transmission approach lies in its inherent scalability and compatibility with future optical network architecture. While the experimental study in this work focused on three-mode fiber transmission, the optimization framework itself and the photonic unitary processor design are fundamentally extendable to systems supporting higher order modes as shown in the simulation. In practical network environments, the photonic node devices should support not only static signal shaping but also switching functionality to dynamically route the optical signal. The MZI-based unitary processor provides a platform for such integration: i.e., the unused ports in the MZI mesh for inter-mode compensation can be repurposed for signal routing, enabling the device to function as both a transmission optimizer and an optical switching engine (see Supplementary Information S8).

While the primary focus of this study is the suppression of DMD, the proposed framework is not limited to a single transmission impairment. The differentiable SDM model can be augmented to target other characteristics such as mode-dependent loss, which is another critical limiting factor in long-haul multi-mode transmission. As shown in Supplementary Information S9, the optimized unitary transformations obtained through this framework also influence MDL. This observation suggests that we can simultaneously not only minimize the digital MIMO cost (e.g., the DMD) but also maximize the transmission capacity (e.g., reducing the mode dependent loss).

It would be worse to compare our approach with conventional mode control techniques such as random mode scrambling and fixed fiber permutations (details are summarized in Supplementary Information S10). While such methods could also decrease the modal delays, their fixed nature limits adaptability. Moreover, finding effective permutations often involves combinatorial search and does not scale well with mode count. As shown in Supplementary S11, optimal permutations vary depending on modal delays, highlighting the limitations of static configurations.

For optimizing the unitary matrix with a given cost function, there are several candidate algorithms besides the gradient-based backpropagation approach. To compare the performance of our demonstrated gradient-based optimization approach, we numerically simulated a possible alternative method called simulated annealing (SA), which is a commonly used technique for exploring high-dimensional solution spaces without gradient information. As a result, we found that the differentiable optimization framework offers distinct advantages in both convergence performance and solution quality (see Supplementary Information S12). This might be because SA typically searches for the optimal matrix by flipping the matrix element from zero to one, which limits the searchable solution space. Also, SA is intrinsically based on the trial-and-error approach requiring many iteration steps for the convergence. On the other hand, the gradient-based methods enable continuous-valued optimization across the full unitary space. As a result, they can discover transformations from the wider solution space. This efficiency provides scalability to larger-scale systems or reconfigurable adaptation and thus reduces the optimization computational cost. It should be noted that SA still retains unique advantages: since it does not rely on gradient information, it remains effective in experimental settings when a differential model is difficult to construct accurately. In such cases, the gradient-free optimization techniques could play a complementary role. Looking ahead, gradient-free methods for physical neural networks such as direct feedback alignment (DFA) [46] or forward-forward learning [47] may allow us to achieve even more robust and adaptable photonic optimization strategies.

**Method**

**Numerical simulation and optimization of parametrized SDM transmission**

The SDM fiber transmission can be modeled by a frequency-dependent transfer matrix $M(\omega)$, which describes the linear transformation applied to input mode amplitudes as they propagate through the fiber:

$$E_{out}(\omega) = M_{total}(\omega)\, E_{in}(\omega), \tag{3}$$

where $E_{in}(\omega)$ and $E_{out}(\omega)$ are the input and output mode amplitude vectors, respectively. $M_{total}(\omega)$ is the end-to-end transmission matrix the transmission after hopping $K$ nodes, which can be decomposed to the following formula [12,34].

$$M_{total}(\omega) = M_{K+1}(\omega) \prod_{k=1}^{K} U_k M_k(\omega), \tag{4}$$

where $M_k(\omega)$ and $U_k$ represents the transmission matrix of the MMF and wavelength-insensitive unitary matrix defined on the photonic processor at the $k$th span ($k = 1, 2, 3, …, K$), respectively. The unitary conversion $U_k$ can be calculated by the following equation.

$$U_k = D' \prod_{(i,j) \in MZI-mesh} T_{ijk}(\xi_1^{(i,j,k)}, \xi_2^{(i,j,k)}), \tag{5}$$

where $i$ and $j$ denote the positions of MZIs in the mesh, $D'$ is a diagonal matrix, and $T_{ijk}$ is the 2×2 unitary conversion [33]. To analyze the effect of mode dispersion, the transmission matrix for an SDM fiber span of length $K$ can be parameterized as:

$$M_k(\omega) = S_k D_k(\omega) V_k^\dagger, \tag{6}$$

where $S_k$ and $V_k$ represents the input and output unitary mode coupling within each fiber span. By considering the transmission matrix near the center frequency $\omega'$, $D_i$ can be described as follows [48]:

$$D_k(\omega) = diag[e^{(\alpha_1 - j(\omega - \omega')\tau_1)}, e^{(\alpha_2 - j(\omega - \omega')\tau_2)}, \ldots, e^{(\alpha_N - j(\omega - \omega')\tau_N)}], \tag{7}$$

where $\alpha_i$ and $\tau_i$ are the mode-dependent gain (or loss) and delay for the $j$th principal mode ($i$ = 1, 2, 3, …, $N$). Considering the phase term the transmission matrix of the transmission matrix, the group delay operator can be defines as follows:

$$G(\omega) = -j \frac{d}{d\omega} log(M_{total}(\omega)) = -j \left(\frac{dM_{total}(\omega)}{d\omega}\right) M_{total}^{-1}(\omega), \tag{8}$$

The term $dM_{total}(\omega)/d\omega$ can be numerically estimated by using following approximation: $dM_{total}(\omega)/d\omega \sim \{M_{total}(\omega + \Delta\omega) - M_{total}(\omega)\}/\Delta\omega$, where $\Delta\omega$ is the frequency differential step. If $M(\omega)$ is unitary, $G(\omega)$ is Hermitian with real eigenvalues $\tau^{total} = [\tau_1^{total}, \tau_2^{total}, \ldots, \tau_N^{total}]$, representing the accumulated modal delays of the different principal modes. By using singular value decomposition (SVD), we can compute $\tau^{total}$ as follows: $S', \tau^{total}, V' = SVD(G)$, where $S'$ and $V'$ corresponds to eigen vectors, and $\tau^{total}$ corresponds to the singular value. As the transmission matrix of multimode fiber $M_k(\omega)$ depends on the time due to the phase fluctuations, the $\tau^{total}$ should be calculated $N$ times to estimate the statistical value. Then, we can estimate maximum differential mode delay, which corresponds to the temporal dispersion length $L$ of the input signal, as follows:

$$L = \frac{1}{N} \sum_{i}^{N} \max_{j} |\tau_j^{total,i} - \tau_1^{total,i}|, \tag{9}$$

where $i$ and $j$ denote the number of simulation trials ($i$=1, 2, …, $N$) and modal number ($j$=1, 2, …, $D$).

As can be seen in Eqs. (1)–(6), all the equations for emulating the statistical DMD are differentiable, including the SVD processing. Therefore, we can optimize the transmission matrix through the controllable unitary matrices $U_k$ by estimating their gradients. Here, we consider the optimization process. To minimize a given cost function $L = \sum_i^N |\tau_i^{total} - \overline{\tau^{total}}|/N$, which corresponds DMD, we can calculate the gradient of $L$ as follows:

$$\frac{\partial L}{\partial U_k} = S'^\dagger \left(\frac{\partial G}{\partial U_k}\right) V' \sim S'^\dagger \frac{1}{\Delta \omega} \left(\frac{\partial M_{total}(\omega+\Delta\omega)}{\partial U_k} - \frac{\partial M_{total}(\omega)}{\partial U_k}\right) V'. \tag{10}$$

By considering the variation $\delta U_k$, we can derive $dM_{total}(\omega)/dU_k$ as follows:

$$\delta M_{total}(\omega) = M_{K+1}(\omega)\left(\prod_{j=k+1}^{K} U_j M_j(\omega)\right) \delta U_k M_k(\omega) \left(\prod_{j=1}^{k} U_j M_j(\omega)\right), \tag{11}$$

$$\frac{dM_{total}(\omega)}{dU_k} = M_{K+1}(\omega)\left(\prod_{j=k+1}^{K} U_j M_j(\omega)\right) M_k(\omega) \left(\prod_{j=1}^{k} U_j M_j(\omega)\right). \tag{12}$$

If we update the unitary matrix $U_k$ using gradient descent based on the gradient $\partial L/U_k$, the $U_k$ will no longer remain a unitary matrix, which introduces unrealistic gains or losses. On the other hand, the parameters of the matrix $U$ can be decomposed to the phase parameters $\xi_1^{(i,j,k)}$ and $\xi_2^{(i,j,k)}$ by the Clements decomposition [33]. Therefore, by computing the following gradient information, we can update the parameters while preserving the unitary constraint on the matrix.

$$\frac{\partial L}{\partial \xi_{1,2}^{(i,j,k)}} = \frac{\partial U_k}{\partial \xi_{1,2}^{(i,j,k)}} \frac{\partial L}{\partial U_k}, \tag{13}$$

The first term can be computed by backward propagation of the Clements mesh [49]. Notably, all the calculations for estimating these gradients can be obtained by using the *autograd* function in the Pytorch [50] module. Note that the optimized unitary matrix $U^{opt}$ is the ideal value neglecting internal error (e.g., the initial phase difference in the MZI arms and coupling ratio error of directional couplers) within the MZI mesh. Therefore, we compensated these errors based on our machine-learning calibration steps by inferencing the internal errors of the MZIs from the measurement data as described below and in Supplementary Information S3.

In practical long-haul transmission, FMF undergoes phase perturbations due to temperature variations and mechanical vibrations, causing $M_{total}(\omega)$ to fluctuate rapidly over time. As a result, direct photonic MIMO processing, such as diagonalizing the MIMO matrix, must dynamically track these variations at high speed, posing significant technical challenges. On the other hand, since DMD can be analyzed based on its statistical time average, even slow or passive circuits can effectively provide substantial improvements.

One difficulty in performing the simulations based on above-described model is how to generate unitary matrices $S_k$ and $V_k$ for the MMF with arbitrary modal crosstalk. While past studies have estimated MMF XT using methods such as wave propagation simulation [36] or random unitary matrix [12,13], generating unitary transformations with varying phase states at each sampling step requires enormous computational resources for backpropagation optimization based on the differentiable model. Therefore, as a more computationally efficient approach, we employed QR decomposition to construct arbitrary unitary matrices. The detailed generation process is provided in Supplementary Information S13.

For the simulations in the three-mode case, the delay between LP01 and LP11 modes was set to 55 ps/km, while $LP_{11a}$ and $LP_{11b}$ modes were assumed to be degenerate with the same delay, which well explains the experimental results in the following section. The strong coupling was assumed within each mode group, while weak inter-group coupling was considered, with negligible mode-dependent loss (MDL). For the ten-mode case, four mode groups were assumed. Based on previous experimental reports, the modal delays were set as [0, 90.4, 90.4, 46.1, 46.1, 46.1, -65.4, -65.4, -65.4, -65.4] ps/km [11]. The installation span $z$ of the photonic processors was set to $z$ = 51.2 km to emulate the experimental condition, excluding the results for span dependency shown in Supplementary Information S1. The transmission step was discretized by $z$/2.

**Fabrication, measurement, and calibration of photonic unitary processor**

The waveguide frontend was fabricated by using silica-based PLC technology with an index contrast of 1.5%. The chip consists of 36 dual-arm MZIs, including eight tapping MZIs for proper calibration. All optical path lengths from the input to output were designed to be equal, eliminating the undesired wavelength dependence. Each MZI comprises two directional couplers (DCs) and four thermo-optic phase shifters based on titanium nitride heaters. Dual-arm operation of MZI was utilized to suppress thermal crosstalk, which limits its maximum phase shift (proportional to the thermal consumption for each heater) below π. The measured total insertion loss was 2.8 dB on average, including the coupling loss between the fiber array and chip and the insertion loss of each MZI (0.157 dB). The implemented matrices on the fabricated MZI mesh and their wavelength dependence were characterized by using the ASE light source and optical spectrum analyzer while the input/output port was scanned using the optical switches. The relationship between the relative phase shift in the MZI and driving current — as well as the static phase error between internal interferometer arms — could be directly calibrated by the measured transmittance while scanning the driving current. However, the actual splitting ratio of DCs ($\alpha^{(i)}$) and the initial phase errors ($\xi_{error}^{(i)}$) at two-external phase shifters remained unknown due to the difficulty of direct measurement, which drastically affected the fidelity between the target and the obtained matrix. Here, we inferred these unknown parameters using a machine-learning technique. We constructed the digital-twin model of our circuit onto the Pytorch-based differential model and trained the unknown parameters (i.e., $\alpha^{(i)}$ and $\xi_{error}^{(i)}$) using the known input and the measured output intensity series (100 random matrices). The detailed calibration steps are described in Supplementary Information S3.

**Transmission experiment**

At the transmitter, 12-GBaud PDM- quadrature-phase-shift-keying (QPSK) signals were generated by LiNbO$_3$ modulators using a binary pattern coded by a low-density parity code (LDPC) with a code rate of 4/5, as defined in the DVB-S2 standard together with a BCH (30832, 30592) code, yielding a normalized generalized mutual information (NGMI) threshold of 0.836. The test channel was subsequently decorrelated using a PDM emulator with a 295-ns delay and then combined with ten other WDM signals that were generated from wavelength selective switch (WSS)-shaped ASE sources with a channel separation of 12.5 GHz. Three-mode MDM-emulated signals were then obtained by splitting these signals through delays lines with delays of 567 and 1204 ns for LP11a and LP11b signals, respectively. The input signals are circulated in the graded-index MMF-based fiber loop via acoustic optical modulators (AOMs), which enables the path switching for the arbitrary timing of signal circulation. The MMF had 51.2-km length, supporting three-mode transmission, including one fundamental mode (LP01) and two degenerated modes (LP11a and LP11b). The setup enables the investigation of the transmission characteristics with varying transmission spans (i.e., the number of the recirculations within the MMF loop). In the MMF loop, the spatial signals were demultiplexed by a mode demultiplexer (DEMUX), and the transmission loss in each span (18 dB) was compensated by erbium-doped fiber amplifiers (EDFAs). The gained signals were input to the photonic unitary processor with the implemented matrix $U$. The output signals were again multiplexed to be spatial signals by a mode multiplexer (MUX). Thus, our experimental setup corresponds to testing the span-independent unitary matrix $U_k = U$ for the three-mode fiber transmission shown in Fig. 1. After transmissions, MDM signals were detected by a coherent SDM receive setup and then processed with a typical offline MIMO-DSP procedure [11], with chromatic dispersion compensation and frequency-domain MIMO equalization performed with an equalizer length up to 800. The experimental total modal dispersion $L$ was quantitatively analyzed by considering the required MIMO equalizer window, which is defined as the time window covering 95% of pulse energy at each distance.

**Data availability**

All the data and methods needed to evaluate the conclusions of this work are presented in the main text and Supplementary Information. The detailed experimental results are available from the corresponding author on reasonable request.

**Code availability**

Codes used in this paper are available from the corresponding author on reasonable request.

**References**

1.  T. Morioka, "New generation optical infrastructure technologies "EXAT initiative" toward 2020 and beyond," Proc. OECC2009, FT4, 2009.
2.  Winzer, P. J., Neilson, D. T. Neilson, and Chraplyvy, A. R. "Fiber-optic transmission and networking: the previous 20 and the next 20 years." *Opti. Exp*. 26, 24190-24239 (2018).
3.  Ellis, A. D., Zhao, J. and Cotter, J. "Approaching the non-linear Shannon limit." *J. Light. Technol*., 28,423–433 (2010).
4.  Sillard, P. et al. "Few-mode fiber technology, deployments, and systems." Proceedings of the IEEE 110, 1804-1820 (2022).
5.  Marom, D. M., Miyamoto, Y., Neilson, D. T., and Tomkos, I. "Optical switching in future fiber-optic networks utilizing spectral and spatial degrees of freedom." Proceedings of the IEEE 110, 1835-1852 (2022).
6.  Puttnam, B. J., Georg, R., and Ruben S. L. "Space-division multiplexing for optical fiber communications." *Optica* 8, 1186-1203 (2021).
7.  Ryf, R. et al. "Mode-division multiplexing over 96 km of few-mode fiber using coherent 6× 6 MIMO processing." *J. Light. Technol.* 30, 521-531 (2012).
8.  Puttnam, B. J. et al. "22.9 Pb/s data-rate by extreme space-wavelength multiplexing." IET Conference Proceedings CP839. Vol. 2023. No. 34. Stevenage, UK: The Institution of Engineering and Technology, 2023.
9.  Shibahara, K. et al. "DMD-unmanaged long-haul SDM transmission over 2500-km 12-core× 3-mode MC-FMF and 6300-km 3-mode FMF employing intermodal interference canceling technique." *J. Light. Technol.* 37, 138-147 (2019).
10. Soma, D. et al. "10.16-Peta-B/s dense SDM/WDM transmission over 6-mode 19-core fiber across the C+ L band." *J. Light. Technol.* 36, 1362-1368 (2018).
11. Shibahara, K., Megumi H., and Yutaka M. "10-spatial-mode 1300-km transmission over 6-LP graded index few-mode fiber with 36-ns modal dispersion." *J Light. Technol.* 42, 1257-1264 (2024).
12. Vijay, A., Krutko, O., Refaee, R., and Kahn, J. M. "Modal statistics in mode-division-multiplexed systems using mode scramblers." *J. Light. Technol*. 43, 845 – 856 (2025).
13. Askarov, D., and Kahn, J. M. "Long-period fiber gratings for mode coupling in mode-division-multiplexing systems." *J. Light. Technol.* 33, 4032-4038 (2016).
14. Gao, X. et al. "Energy-efficient hybrid analog and digital precoding for mmWave MIMO systems with large antenna arrays." *IEEE J. Sel. Areas Commun.* 34, 998-1009 (2016).
15. Wang, P., Fang, J., Yuan, X., Chen, Z., and Li, H. "Intelligent reflecting surface-assisted millimeter

wave communications: Joint active and passive precoding design." *IEEE Trans. Veh. Technol.* 69, 14960-14973 (2020).

16. Shallah, A. B. et al. "Recent developments of butler matrix from components design evolution to system integration for 5g beamforming applications: A survey." *IEEE Access* 10, 88434-88456 (2022).
17. Zhang, W., Tait, A., Huang, C. et al. "Broadband physical layer cognitive radio with an integrated photonic processor for blind source separation." *Nat. Commun.* 14, 1107 (2023). https://doi.org/10.1038/s41467-023-36814-4
18. Pérez-López, D., Gutierrez, A., Sánchez, D. et al. "General-purpose programmable photonic processor for advanced radiofrequency applications." *Nat. Commun.* 15, 1563 (2024). https://doi.org/10.1038/s41467-024-45888-7
19. Sanjari, P. and Aflatouni, F. "An integrated photonic-assisted phased array transmitter for direct fiber to mm-wave links." *Nat. Commun.* 14, 1414 (2023). https://doi.org/10.1038/s41467-023-37103-w
20. Pablo, M-C. et al. "Ultrabroadband high-resolution silicon RF-photonic beamformer." *Nat. Commun*. 15.1 (2024): 1433.
21. Shen, Y. et al. "Deep learning with coherent nanophotonic circuits." *Nat. Photon.* 11, 441–446 (2017).
22. Shastri, B. J. et al., "Photonics for artificial intelligence and neuromorphic computing." *Nat. Photon.* 15, pp. 102–114 (2021).
23. Feldmann, J. et al. "Parallel convolution processing using an integrated photonic tensor core." Nature 589, 52–58 (2021).
24. Harris, N. C. et al. "Linear programmable nanophotonic processors." *Optica* **5**, 1623–1631 (2018).
25. Xu, Z. et al. "Large-scale photonic chiplet Taichi empowers 160-TOPS/W artificial general intelligence." *Science* 384, 202-209 (2024).
26. Dong, M. et al. "Programmable photonic integrated meshes for modular generation of optical entanglement links." npj Quantum Inf 9, 42 (2023).
27. Ikeda, K. et al. "16× 16 MZI-based photonic accelerator toward space and wavelength division multiplexing." SPIE Emerging Topics in Artificial Intelligence (ETAI), (2024).
28. Kim, K., Park, K., Park, H., Yu, S., Park, N., & Piao, X. "Programmable photonic unitary circuits for light computing." *Nanophotonics*, (2025): https://doi.org/10.1515/nanoph-2024-0602.
29. Tanomura, R., et al. "All-optical mimo demultiplexing using silicon-photonic dual-polarization optical unitary processor." *J. Light. Technol.* 41.12 (2023): 3791-3796.
30. SeyedinNavadeh, S. et al. Determining the optimal communication channels of arbitrary optical systems using integrated photonic processors. *Nat. Photon.* 18, 149–155 (2024).
31. Takahashi, H. "High performance planar lightwave circuit devices for large capacity transmission."

*Opt. Exp.* 19, B173 (2011).

32. Nakajima, M., Tanaka, K. & Hashimoto, T. "Scalable reservoir computing on coherent linear photonic processor." *Commun. Phys.* 4, 1–12 (2021).
33. Clements, W. R., Humphreys, P. C., Metcalf, B. J., Kolthammer, W. S., and Walmsley, I. A. "Optimal design for universal multiport interferometers." *Optica* 3, 1460-1465 (2016).
34. Ho, K. P., and Kahn, J. M., "Mode-dependent loss and gain: statistics and effect on mode-division multiplexing." *Opt. Exp.* 19, 16612-16635 (2011)
35. Ho, K. P. and Kahn, J. M. "Statistics of group delays in multimode fiber with strong mode coupling." *J. Light. Technol.* 29, 3119-3128 (2012).
36. Giammarco, D. S. et al. "Enhancing Long-Haul 15-Mode Fiber Performance: Mode Permutation for Reduced Modal Dispersion." *J. Light. Technol.* 43, 481-491 (2024).
37. Sakamoto, J., Goh, T., Katayose, S., Kasahara, R. & Hashimoto, T. "Shape-optimized multi-mode interference for a wideband visible light coupler." *Opt. Commun.* 443, 221 (2019).
38. Carolan, J. et al. "Universal linear optics." *Science* 349, 711 (2015).
39. Aratake, A. "High reliability of silica-based 1 × 8 optical splitter modules for outside plant." *J. Light. Technol.* 34, 27 (2016).
40. Akatsuka, T. et al. "Optical frequency distribution using laser repeater stations with planar lightwave circuits." *Opt. Exp.* 28, 9186-9197 (2020).
41. Ikeda, K., Nakajima, M., Kita, S., Shinya, A., Notomi, M., and Hashimoto, T. "High-fidelity WDM-compatible photonic processor for matrix-matrix multiplication." CLEO, JTh2A-87 (2024).
42. Tanizawa, K., Suzuki, K., Ikeda, K., Namiki, S., and Kawashima, H. "Non-duplicate polarization-diversity 8× 8 Si-wire PILOSS switch integrated with polarization splitter-rotators." *Opt. Exp.* 25, 10885-10892 (2016).
43. Shibahara, K. et al., "Long-haul three-mode-multiplexed transmission employing on-chip mode permutation technique based on a programable photonic unitary processor", OECC, 194 (2024).
44. Cho, J., Schmalen, L., and Winzer, P. J. "Normalized generalized mutual information as a forward error correction threshold for probabilistically shaped QAM." 2017 European Conference on Optical Communication (ECOC). DOI: 10.1109/ECOC.2017.8345872.
45. Shibahara, K., Mizuno, T., Lee, D., Miyamoto, Y., Ono, H., Nakajima, K.,Saitoh, S., Takenaga, and K., Saitoh, K. "DMD-Unmanaged Long-Haul SDM Transmission Over 2500-km 12-core × 3-mode MC-FMF and 6300-km 3-mode FMF Employing Intermodal Interference Cancelling Technique." Optical Fiber Communication Conference, Th4C.6 (2018).
46. Nakajima, M., Inoue, K., Tanaka, K., Kuniyoshi, Y., Hashimoto, T., and Nakajima, K. "Physical deep learning with biologically inspired training method: gradient-free approach for physical hardware." *Nat. Commun.* 13, 7847 (2022).
47. Hinton, G. "The forward-forward algorithm: Some preliminary investigations." arXiv:2212.13345

(2022).

48. Barbosa, F. A. and Ferreira, F. M. "On a scalable path for multimode SDM transmission. *J. Light. Technol*." 42, 219-228 (2023).

49. Jing, L. et al. "Tunable efficient unitary neural networks (eunn) and their application to rnns." 34th International Conference on Machine Learning (ICML), 1733-1741 (2017).

50. Paszke, A. et al. "PyTorch: An Imperative Style, High-Performance Deep Learning Library." Advances in Neural Information Processing Systems 32, 8024–8035 (2019).

**Acknowledgement**

This study. was partially supported by NICT, Japan, under commissioned research JP012368C01001. We are grateful to Drs. H. Sawada and K. Sawada for their contributions to the calibration algorithm of the MZI mesh. We thank F. Sugimoto for technical contributions for the optical experiment. We also thank S. Kita for the fruitful discussions.

**Contributions**

M. Nakajima and K.S. conceived the basic concept of the presented parametrized SDM transmission. M. Nakajima and K.I. developed the differentiable simulator and its optimizer with technical support from K.S. and A.K. M. Nakajima and K.S. performed the numerical simulation by using the constructed simulator. M. Nakajima and K.I. designed, fabricated, and characterized the programmable photonic unitary circuit. K.S. and A.K. constructed the experimental setup for the long-haul SDM transmission. T.K., Y.M., M. Notomi, and T.H. supervised the project. M. Nakajima wrote the initial draft of the manuscript. All the authors discussed the results and contributed to writing the manuscript.

**Competing interests**

The authors have filed patent application no. PCT/JP2024/023068 on parametrized SDM node systems.

**Figure captions**

**Figure 1. Parameterized differentiable SDM transmission with photonic unitary processors.** *(a) Schematic explanation of parametrized SDM transmission. Statistical characteristics can be optimized by using trainable photonic unitary processors enabling direct optimization of the transmission matrix to reduce the pre/post MIMO processing load. (b) Optimization of photonic unitary conversion based on differential propagation model. (c) Concrete step of optimization: trainable phase parameters ξ in the photonic unitary processors are optimized by estimating the gradient ∂L/∂ξ.*

**Figure 2. Optimization of unitary matrix to reduce modal dispersion of MMF.** *Convergence process of modal dispersion over training epochs for (a) three and (b) ten- mode fibers under different coupling conditions of the MMF. Modal dispersion as a function of transmission distance for (c)–(f) three and (g)–(j) ten- mode fibers before (i.e., random matrixes drawn as blue lines) and after (red line) optimization of unitary matrices with four possible MMF coupling scenario; (c), (g) no coupling, (d), (h) strong coupling within degenerated modes, (e), (i) strong coupling within degenerated modes and weak coupling with inter mode-group, and (f), (j) strong coupling within all modes. The upper part of each figure shows defined MMF transmission matrices and optimized unitary matrices. All filled areas denote standard deviations based on 100 simulation trials with changing random seeds for generation phase perturbation and MMF matrix.*

**Figure 3. PLC-based photonic unitary processor for SDM-WDM-PDM transmission.** *(a) Schematic of photonic unitary processors required for parametrized SDM transmission. The photonic processor should execute unitary operation insensitive to the wavelength and polarization state, as the standard SDM transmission utilizes not only space division but also wavelength and polarization division. (b) Photograph of fabricated PLC-based Clements-type photonic unitary circuit (50 mm × 25 mm). (c) Fiber-to-fiber insertion loss as a function of wavelength. Color difference corresponds to difference of input/output ports.. (d) Fidelity histogram at 1550 nm before and after calibration (inset: learning convergence curve). e), (f) Target and measured transmission matrix at 1530, 1550, and 1570 nm for (e) identity and (f) random unitary matrix. (g) Wavelength dependence of fidelity, comparing individually optimized (red) and optimized at 1550 nm (blue). The inset shows the estimated splitting ratio of the directional coupler in the MZI.*

**Figure 4.** *Experimental setup and basic characteristics of parameterized SDM transmission. (a) Schematic diagram of the experimental setup for long-haul three-mode fiber transmission, incorporating a photonic unitary processor. (b) Comparison of simulated and experimental mode dispersion as a function of Θ. (e) Implemented unitary matrices for different Θ setups. θ. (c) Numerical*

*simulation and (d) experimental results of mode delay as a function of transmission distance and unitary transformation parameter Θ. (e) Mode delay accumulation over transmission distance for different Θ (Θ=0, -π, -2π/3), comparing experimental (dot plots) and simulation (slid lines) results. In the transmission experiment, we sent 33,360 symbols and measured the impulse response through digital MIMO postprocessing and defined its 95% region as the differential mode delay. The simulations were repeated 100 times, and averaged values are plotted. The shaded region in (e) corresponds to the standard deviation of the simulation.*

**Figure 5. Optimization using differentiable SDM transmission model.** *(a), (b) Potential landscape of (a) simulated and (b) experimental mode delay as a function of trainable parameters. Solid lines show the trajectory under the optimization step landscape for different initial states. (c) Training curve of modal dispersion for simulation (solid lines) and experiment (circle and triangle plot). The simulations were repeated 100 times, and averaged values are plotted. In the transmission experiment, we sent 33,360 symbols and measured the impulse response through digital MIMO postprocessing and defined its 95% region as the differential mode delay. The shaded region in (c) corresponds to the standard deviation of the simulation.*

**Figure 6. Transmission characteristics of parameterized SDM transmission with optimized photonic unitary processing.** *(a) Mode delay accumulation over transmission distance for different before (identity matrix) and after optimization, comparing experimental (dot plots) and simulation (slid lines) results. For the experimental results, the results for wavelengths of 1541.349-, 1550.057-, and 1558.173-nm are also plotted to confirm the adaptivity to the WDM signals. (b), (c) evolution of Impulse response along transmission distance for (b) 1541.349 and (c) 1558.173 nm. (d) Received constellation after 1024-km transmission for each spatial and polarization modes. The estimated NGMI values are shown in the lower part. (e) NGMI as a function of transmission distance. The dashed line indicates error-free threshold of 0.836. In the transmission experiment, we sent $2^{XXX}$ symbols and measured the impulse response through digital MIMO postprocessing and defined its 95% region as the differential mode delay. The simulations were repeated 100 times, and averaged values are plotted. The shaded region in (c) corresponds to the standard deviation of the simulation.*

# Supplementary Information for Programmable Photonic Unitary Processor Enables Parametrized Differentiable Long-Haul Spatial Division Multiplexed Transmission

Mitsumasa Nakajima[1,a,*], Kohki Shibahara[2,*], Kohei Ikeda[3,4,*], Akira Kawai[2], Masaya Notomi[3,4], Yutaka Miyamoto[2], and Toshikazu Hashimoto[1]

[1]NTT Device Technology Laboratories, 3-1 Morinosato-Wakamiya, Atsugi, Kanagwa, Japan

[2]NTT Network Innovation Laboratories, 1-1 Hikarinooka, Yokosuka, Kanagwa, Japan

[3]NTT Basic Research Laboratories, 3-1 Morinosato-Wakamiya, Atsugi, Kanagwa, Japan

[4]NTT Nanophotonics Center, 3-1 Morinosato-Wakamiya, Atsugi, Kanagwa, Japan

b. mitsumasa.nakajima@ntt.com

* Equally contributed to this paper

**S1. Span dependency of optimized mode dispersion**

To assess the influence of unitary processor spacing on the suppression of modal dispersion, we evaluated the performance of the proposed parameterized SDM transmission framework under different span lengths between unitary processing nodes. Figure S1(a) shows the accumulated mode dispersion as a function of total transmission distance for four different span lengths: 204.8 (brown), 51.2 (red), 25.6 (blue), and 10.24 km (green) with the total dispersion length of 1331.2 km maintained. The results clearly indicate that shorter spans, corresponding to more frequent deployment of unitary processors, enable stronger suppression of differential mode delay (DMD). Figure S1 (b) summarizes the trade-off between mode dispersion and hardware complexity. The blue curve plots the final DMD (after 1331.2 km) versus the span length. The red curve (right axis) indicates the number of required photonic unitary processors as a function of span length. These results highlight a clear trade-off: while dense compensation (shorter spans) is more effective for DMD reduction, it increases hardware complexity and deployment cost. In practical networks, this trade-off must be balanced based on existing node spacing, amplifier positioning, or ROADM intervals. In our experimental setup, a span length of 51.2 km was selected to reflect typical transmission segment lengths and constraints imposed by laboratory fiber availability.

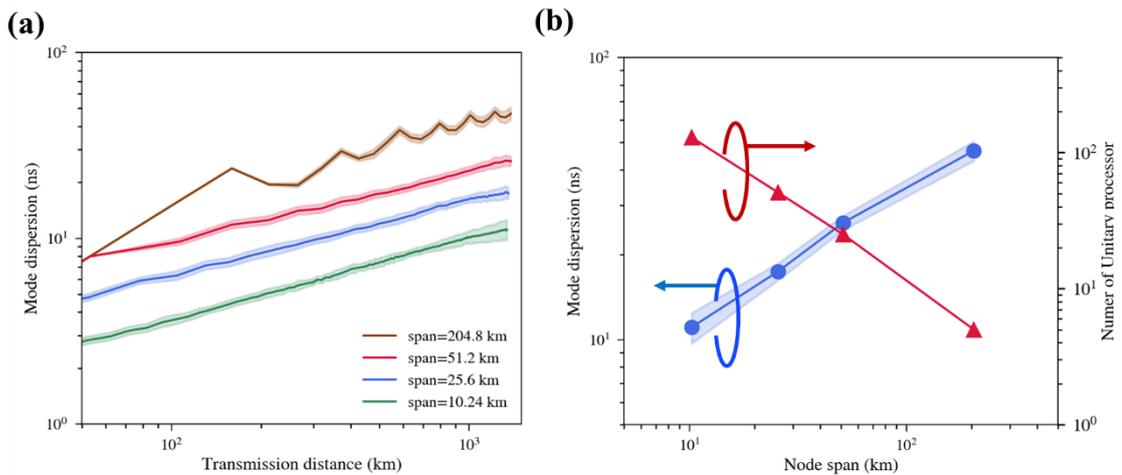

*Figure S1. Span length dependency of differential mode delay.* (a) Optimized mode delay accumulation over transmission distance with different span length for 204.8 (brown), 51.2 (red), 25.6 (blue), and 10.24 km (green) with the total dispersion length of 1331.2 km maintained. (b) Ttrade-off between mode dispersion and hardware complexity. The blue curve plots the final DMD (after 1331.2 km) versus the span length. The red curve (right axis) indicates the number of required photonic unitary processors as a function of span length.

**S2. Comparison of silica- and silicon-based photonic unitary processors**

Table S1 compares PLC technology with silicon photonics. (SiPh) platforms for implementing the broadband high-fidelity unitary processors required for parametrized SDM transmission. Silicon photonics enables highly compact integration but suffers from significant fabrication-induced phase errors, which causes lower baseline fidelity of the unitary transformation and wavelength-dependency even with a path-independent Clements configuration. Additionally, SiPh exhibits intrinsic polarization dependence, necessitating polarization diversity schemes [42]. In contrast, PLC-based waveguides, fabricated from silica—the same material as standard optical fibers—offer inherently broadband and polarization-independent characteristics. The PLC-based circuits have wider fabrication tolerance thanks to the moderate order of the refractive index difference between the core and clad, suppressing significant fabrication errors, which are impossible to compensate.

|  | Si photonics | | Silica-based photonics (PLC) | |
| --- | --- | --- | --- | --- |
|  | Reck | Clements | Reck | Clements (Ours) |
| **Fidelity** | ☺ Medium | ☺ Medium | ☺ High | ☺ High |
| **Wavelength dependency** | ☹ High (due to path difference) | ☺ Medium (due to MZI) | High (due to path difference) | ☺ Low (over C-band) |
| **Polarization dependency** | ☹ High (due to Si property) | ☹ High (due to Si property) | ☺ Low (thanks to silica property) | ☺ Low (thanks to silica property) |
| **Calibration** | ☺ Easy | ☹ Difficult | ☺ Easy | ☹ Difficult |

*Table S1. Comparison of silicon and silica-based photonic unitary processors.*

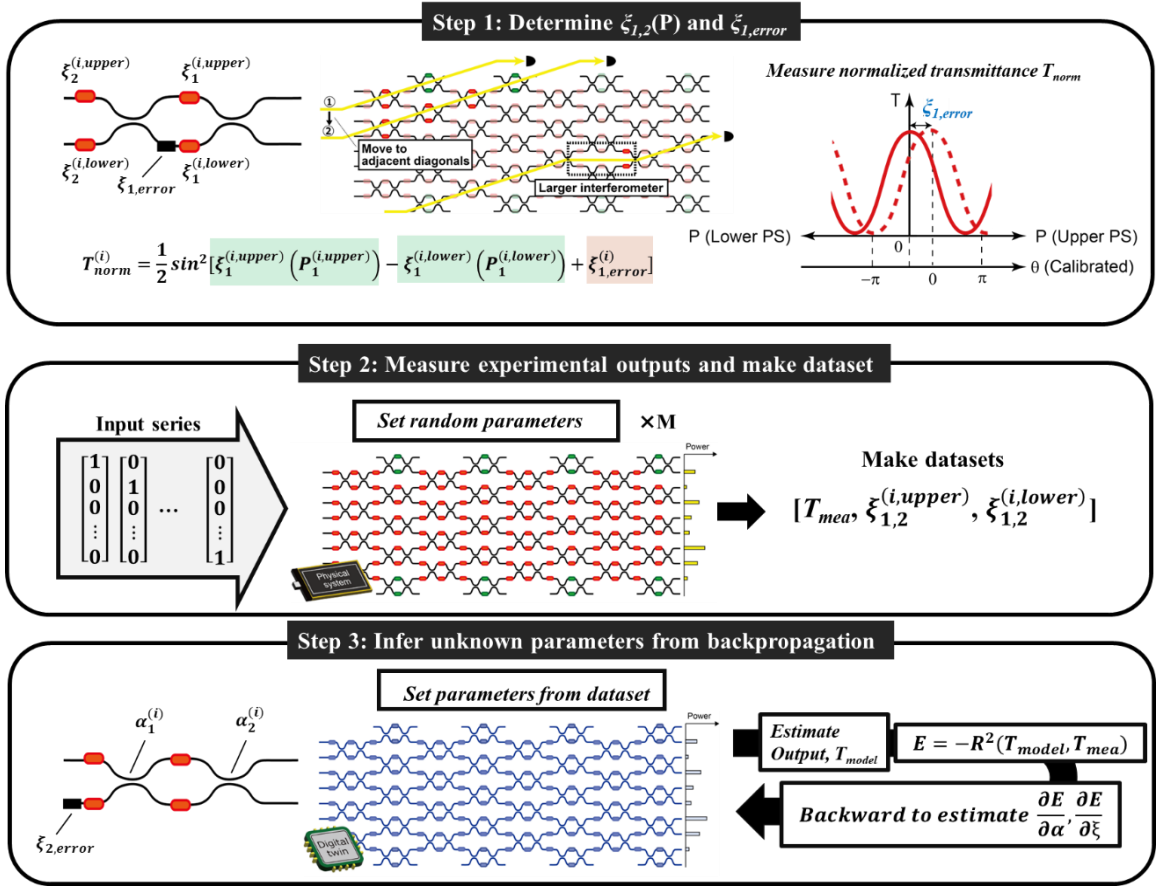

*Figure S2. Calibration method for photonic unitary processor. In step 1 (top), parameters for internal phase shifters were determined. In step 2, training data for model-based inference were acquired. In step 3, remaining parameters were inferred from obtained data.*

**S3. Calibration method for photonic unitary processor**

To ensure high-fidelity transformation in our MZI-based unitary photonic processor, we developed a three-step calibration method that estimates and compensates for unknown phase errors originating from fabrication variations in the optical waveguides. The procedure is summarized in Figure S2.

**Step 1: Determine the parameters for phase shifters**

Each MZI cell contains four thermo-optic phase shifters: two in the internal arms (upper and lower arms) and two in the external arms (upper and lower). In step 1, we first determined the parameters for the internal phase shifters (denoted as 1). Due to fabrication imperfections, these exhibit the static phase offsets between the two interferometer arms denoted by $\xi_{1,error}^{(i)}$ superimposed on the controlled phase shifts $\xi_1^{(i,upper)}(P_1^{(i,upper)})$ and $\xi_1^{(i,lower)}\left(P_1^{(i,lower)}\right)$, where $P$ is the dissipated power in the phase shifter and upper and lower indicate the phase shifter at the upper and lower in $i$th MZI arm. As illustrated in Fig. S2 (top), we first swept the $P_1^{(i,upper)}$ and $P_1^{(i,lower)}$ for each phase shifter independently while monitoring the output transmittance at cross port. The normalized concatenated

optical transmittance $T_{norm}$ follows a $\sin^2$ dependence as follows:

$$T_{norm}^{(i)} = \frac{1}{2}\sin^2[\xi_1^{(i,upper)}\left(P_1^{(i, upper)}\right) - \xi_1^{(i,lowwer)}\left(P_1^{(i, lowwer)}\right) + \xi_{1,error}^{(i)}], \qquad (S1)$$

By fitting the observed output transmission, we extract the static phase offset $\xi_{1,error}^{(i)}$ and the modulation efficiencies of the phase shifters, storing them as the correction parameters. The parameters for the external phase shifters (denoted as 2) are also determined with a similar procedure. To calibrate the external phase shifters, we construct a larger interferometer by combining neighboring MZI cells, effectively placing the external phase shifters inside the resulting MZI structure to allow interferometric measurement. We note that while the modulation efficiencies of the external phase shifters are obtained through this procedure, determining their individual static phase offsets $\xi_{2,error}^{(i)}$ is challenging. Therefore, these parameters are inferred using machine learning in Step.3.

**Step 2: Acquiring training data using random parameter series**
After individually identifying the static offsets of the internal phase shifters and the modulation efficiencies for all phase shifters, we generate 100 random matrix configurations by applying random values to the phase shifters. For each random matrix configuration, we sequentially inject light into each input port using a fiber switch, while the output powers at all output ports are measured to obtain the corresponding output power vectors $T_{mea}$, building a dataset [$T_{mea}$, $\xi_{1,2}^{(i,upper)}$, $\xi_{1,2}^{(i,lowwer)}$] for supervised learning.

**Step 3: Backpropagation-based parameter refinement**
In the final step, we employ a model-driven optimization procedure to refine the remaining unknowns in the optical model, particularly the static phase offsets of the external phase shifters $\xi_{2,error}^{(i)}$ and splitting ratios of the directional couplers in each MZI ($\alpha^{(i)}$). We constructed the differentiable digital twin of the fabricated MZI considering above-described phase and splitting ratio error. Using the dataset generated in Step 2, we define a loss function based on the $R^2$ value between the measured and predicted transmittances $T_{mea}$ and $T_{model}$:

$$E = -R^2(T_{model}, T_{mea}). \qquad (S2)$$

We then apply backpropagation through the differentiable photonic circuit model to estimate the gradients with respect to the residual error parameters $\alpha^{(i)}$ and $\xi_{2,error}^{(i)}$, as shown in Fig. S2 (bottom).

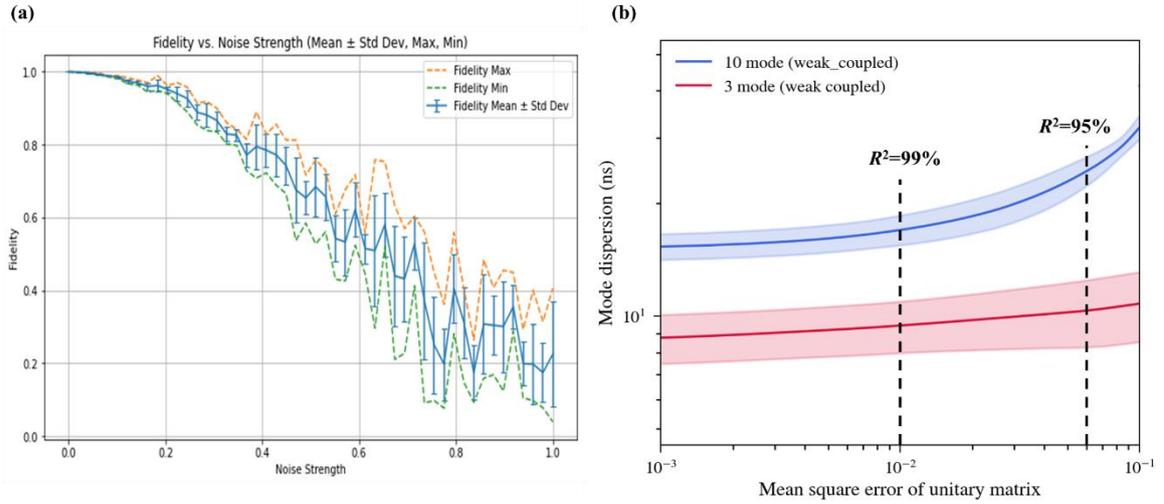

*Figure S3. Impact of error of unitary processor on optimized modal dispersion. (a) Fidelity of the unitary matrix as a function of noise strength. (b)Differential mode delay as a function of mean square error between ideal and non-ideal (noise introduced) photonic unitary processor for three (red) and ten (blue) mode fiber cases.*

(bottom).

## S4. Modal dispersion degradation caused by errors in photonic processor

To examine how deviations from the ideal unitary transformation affect the differential mode delay, we devised a method to introduce controlled perturbations to unitary matrices while strictly preserving their unitarity. In this approach, we begin with an optimized unitary matrix $U^{opt}$, which can be estimated by the gradient-based method described in the main article. Next, we generate the random unitary matrix $U^{rand}$. The $U^{rand}$ can be expressed as $U^{rand} = \exp(jH^{rand})$, where $H^{rand}$ is a Hermitian matrix. To control the degree of deviation from the original matrix, we scale the Hermitian generator by a real coefficient $n$, producing a new Hermitian matrix $H'=nH^{rand}$. Because Hermitian matrices remain Hermitian under real scalar multiplication, $H'$ also satisfies the Hermitian condition. We then construct a new perturbed unitary matrix $U' = \exp(jH')$, which remains strictly unitary by the definition of the matrix exponential. As $n$ increases from zero, the matrix gradually diverges from the original matrix, introducing a tunable level of transformation error. To introduce statistical variation, a new random unitary matrix is resampled in each realization, allowing the evaluation of DMD performance over an ensemble of noisy but valid unitary transformations. Figure S3(a) summarizes the resulting fidelity distribution as a function of noise strength $n$. The average fidelity decreases monotonically with increasing noise, suggesting successful operation of our noise emulating approach. Figure S3(b) further illustrates how DMD degrades as a function of the mean square error (MSE) between the ideal and perturbed unitary matrices for three-mode (red) and ten-mode (blue) cases. The results confirm that for $R^2$ levels above 0.95, the degradation in DMD remains limited, even in the more sensitive ten-mode scenario. These findings justify the experimental implementation of our photonic processors with measured fidelities around 96% across the C-band. The fidelity tolerance demonstrated here ensures that the proposed approach is robust to practical imperfections arising from fabrication and calibration errors.

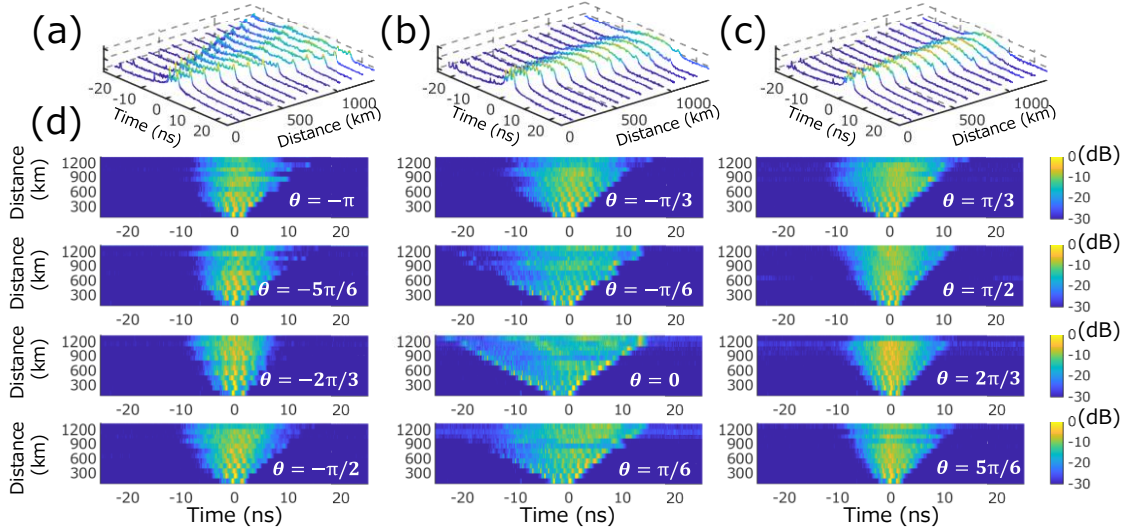

***Figure S4. Θ dependency of impulse response evolution.*** *(a)-(c) Three-dimensional views of impulse response evolutions in (a) weak coupling with Θ=0, (b) intermediate mixing with Θ=π/3, and (c) strong coupling via on-chip mode permutation with Θ=2π/3. (d) Impulse response growth along distance for the whole set of Θ.*

## S5. Distance and Θ dependency of modal dispersion

Figure S4(a)–(c) depict the three-dimensional evolution of impulse response magnitude in dB units along transmission distances, with the rotation angle *Θ* varied within the set of {0,π/3,2π/3}, respectively. M with *Θ*=0 [Fig. S4(a)] yields an identity matrix, resulting in a weakly coupled transmission regime where DMD accumulated nearly linearly with distance, leading to pronounced pulse spreading. Conversely, altering *Θ*=π/3 [Fig.S4(b)] or *Θ*=2π/3 [Fig. S4(c)] as a mixing or permutation matrix, respectively, modifies the behavior of DMD accumulation with a notable reduced pulse spreading. Figure S4(d) highlights how pulse spreading grows with the entire set of θ in the range of [-π,π). Sub-linear pulse broadening was observed within the regime of |*Θ*|<~π/6. However, it became evident that pulse energy was relatively confined within ±10 ns as *Θ* approached ±2π/3.

## S6. Dependence of optimization uunitary matrix on initial parameters

In gradient-based optimization of high-dimensional unitary transformations, the final solution can depend strongly on the initial conditions due to the presence of multiple local minima in the cost function landscape. To investigate this effect, we performed multiple optimization trials starting from different random initial unitary matrices and compared the resulting optimized solutions. Figure S5 shows four examples of optimized 3×3 unitary matrices obtained from distinct initial conditions.

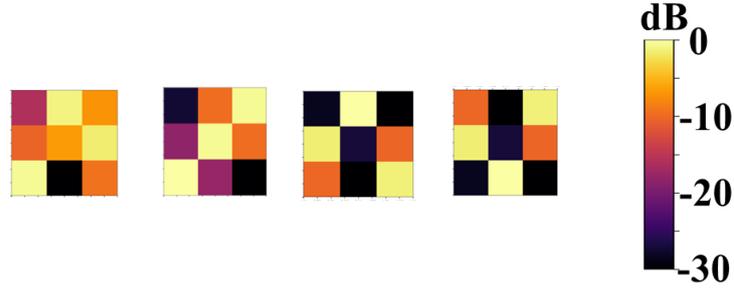

*Figure S5. Dependence of optimized unitary matrix on initial parameters.* Intensity of optimized unitary matrices with four different initial parameters. Obtained matrices are highly dependent on initial condition.

Although all transformations achieve similar performance in terms of minimizing modal dispersion, their matrix structures differ significantly. This suggests the existence of multiple locally optimal configurations in the solution space that are nearly equivalent in performance but not identical in physical transformation. Each matrix is visualized as a heatmap of the absolute values of its elements, revealing differing modal coupling patterns. For instance, some solutions exhibit strong diagonal elements corresponding to near-permutation behavior, while others feature more distributed coupling among modes.

## S7. Dependence of differential mode delay on angles in rotation matrix

Through the simulation in Supplementary Information S4, we observed that different optimal unitary matrices appeared depending on the initial conditions of each trial. This suggests that multiple solutions exist due to the nature of the optimization landscape. To better visualize how the optimization parameters evolve within the unitary matrix, we introduced the following 3D rotation matrix with three independent rotation parameters, $\theta$, $\varphi$, and $\psi$, to simplify the optical circuit parameters:

$$U_{rot}(\theta, \varphi, \psi) = \begin{pmatrix} \cos(\theta)\cos(\varphi) & -\sin(\theta)\cos(\varphi) + \cos(\theta)\sin(\varphi)\sin(\psi) & \sin(\theta)\sin(\varphi) + \cos(\theta)\sin(\varphi)\cos(\psi) \\ \sin(\theta)\cos(\varphi) & \cos(\theta)\cos(\varphi) + \sin(\theta)\sin(\varphi)\sin(\psi) & -\cos(\theta)\sin(\varphi) + \sin(\theta)\sin(\varphi)\cos(\psi) \\ -\sin(\varphi) & \cos(\varphi)\sin(\psi) & \sin(\varphi)\sin(\psi) \end{pmatrix}$$

. (S3)

To evaluate the sensitivity of the optimized unitary transformation to the auxiliary phase parameter $\theta$, $\varphi$, and $\psi$, we conducted a series of simulations in which $\theta$, $\varphi$, and $\psi$ were systematically varied while the other parameters in the transformation were kept constant. Figure S6 displays the differential mode delay as a function of $\varphi$ and $\psi$ for different values of $\theta$. Despite the variation in $\theta$, the overall structure of the matrices remains highly consistent, indicating that the optimization landscape is effectively flat along the $\theta$ axis. This suggests that $\theta$ plays a negligible role in determining the cost function value, particularly in the context of minimizing differential mode delay. This behavior supports the simplification adopted in the main text, where the unitary transformation was parameterized using only the two dominant variables, $\phi$ and $\psi$.

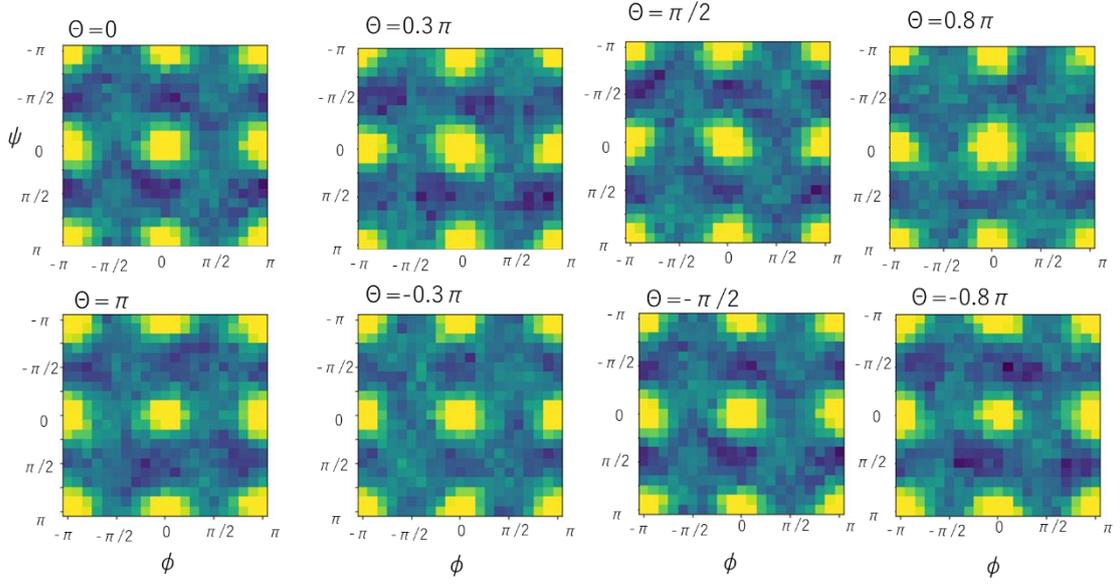

*Figure S6. Dependence of differential mode delay on angles in rotation matrix.* Differential mode delay as a function of $\varphi$ and $\psi$ for different values of $\theta$

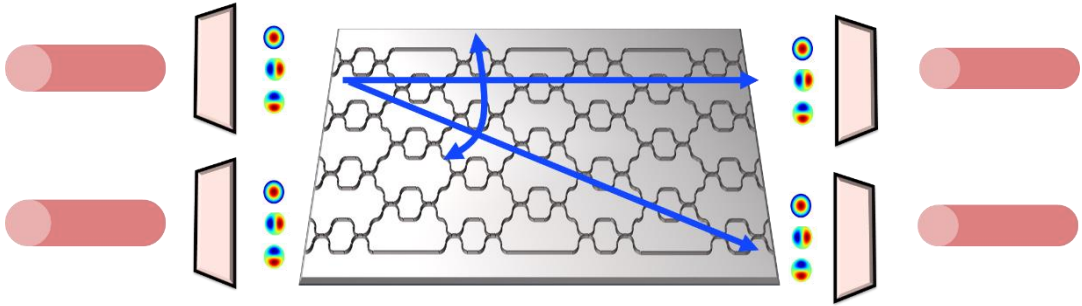

*Figure S7. Integration of switching and modal control function.* Schematic illustration of integrated photonic node enabling modal control described in this paper and optical switching function. It can be realized on programmable photonic unitary processor.

## S8. Adaptivity to future reconfigurable SDM nodes

In addition to its role in in-line modal dispersion optimization, the photonic unitary processor based on a Clements-style MZI mesh architecture can also function as a spatial-mode switching device. This dual functionality makes it highly compatible with reconfigurable photonic network nodes such as ROADMs and optical cross-connects in SDM-WDM networks. As illustrated in Fig. S7, the processor can perform arbitrary unitary transformations, including mode permutation and selection operations. This enables dynamic routing of spatial modes between input and output ports. For instance, by configuring the internal phase shifters appropriately, specific input modes can be directed to arbitrary output modes, allowing the processor to operate as a low-loss, broadband mode switch.

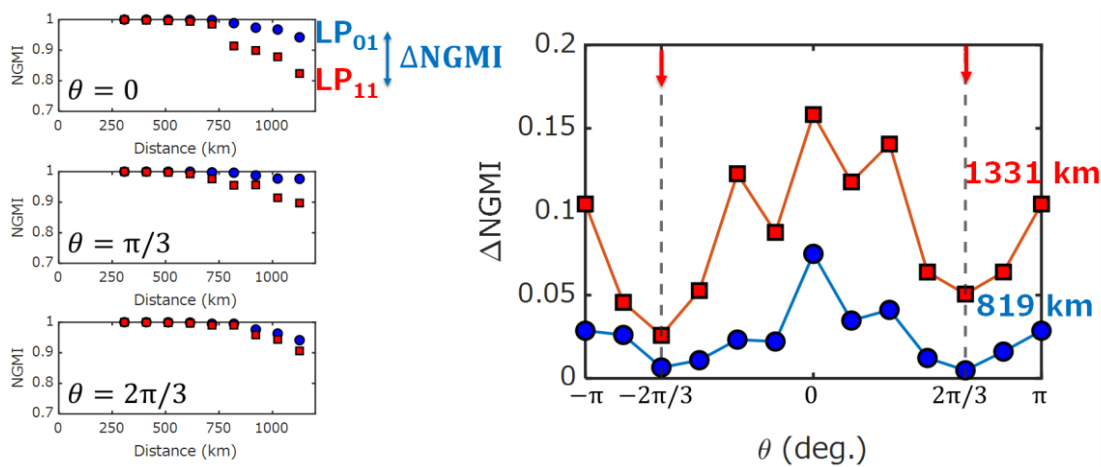

*Figure S8. Controllability of mode dependent loss using photonic unitary processor.* (a) NGMI performance along distance with $\theta \in \{0, \pi/6, 2\pi/3\}$. (b) NGMI difference between LP01 and LP11a/11b at 819 and 1331 km as a function of $\theta$. Vertical broken lines are drawn for the cases of on-chip mode permutation with $\theta = \pm 2\pi/3$

Such flexibility is particularly advantageous for node-level applications, where it is often necessary to reshape, reroute, or drop specific spatial modes. Importantly, because the same Clements MZI mesh structure is used for both unitary optimization and switching, the processor can fulfill both roles within a unified hardware platform without architectural changes. Furthermore, the silica-based PLC implementation of the processor offers low insertion loss, wide spectral flatness, and polarization insensitivity, all of which are critical for ensuring robust performance in real-world network deployments. These characteristics, combined with its programmability and integrability, position the processor as a promising candidate for future SDM node devices that demand both physical-layer signal shaping and reconfigurable connectivity.

## S9. Controlability of mode dependent loss uisng photonic unitary processor

To investigate the possibility for optimizing another transmission characteristic besides the modal delay, we focused on understanding the impact of mode-dependent loss (MDL). Instead of estimating MDL directly, we utilized the difference in NGMI between the fundamental mode (LP01) and higher-order modes (LP11a/11b), because of potential estimation errors of direct MDL estimation from MIMO-DSP equalizer taps [S1]. Figure S8(a) depicts mode-dependent NGMI as a function of distance for different values of $\Theta$ [from eq. (1) in the main article] within the set of $\theta \in \{0, \pi/3, 2\pi/3\}$, indicating that the performance of the MDM signal became almost identical as the regimes approached mode permutation. Figure S8(b) illustrates the variation of the NGMI difference between LP01 and LP11a/11b, denoted as ΔNGMI, at distances of 819 and 1331 km as θ was varied. The smallest ΔNGMIs were attained when mode permutation regimes were achieved. Based on these findings, at least for a given unitary matrix represented in eq. (1) in the main article, we can infer that mode permutation is the optimal MDM transmission strategy for mitigating impact both of DMD and MDL. We confirmed that NGMI with certain sets of $\Theta$ exceeded the NGMI threshold of 0.836 at this distance. This suggests the fundamental applicability of our proposed parametrised SDM scheme is effective for not only the DMD but also the MLD optimization.

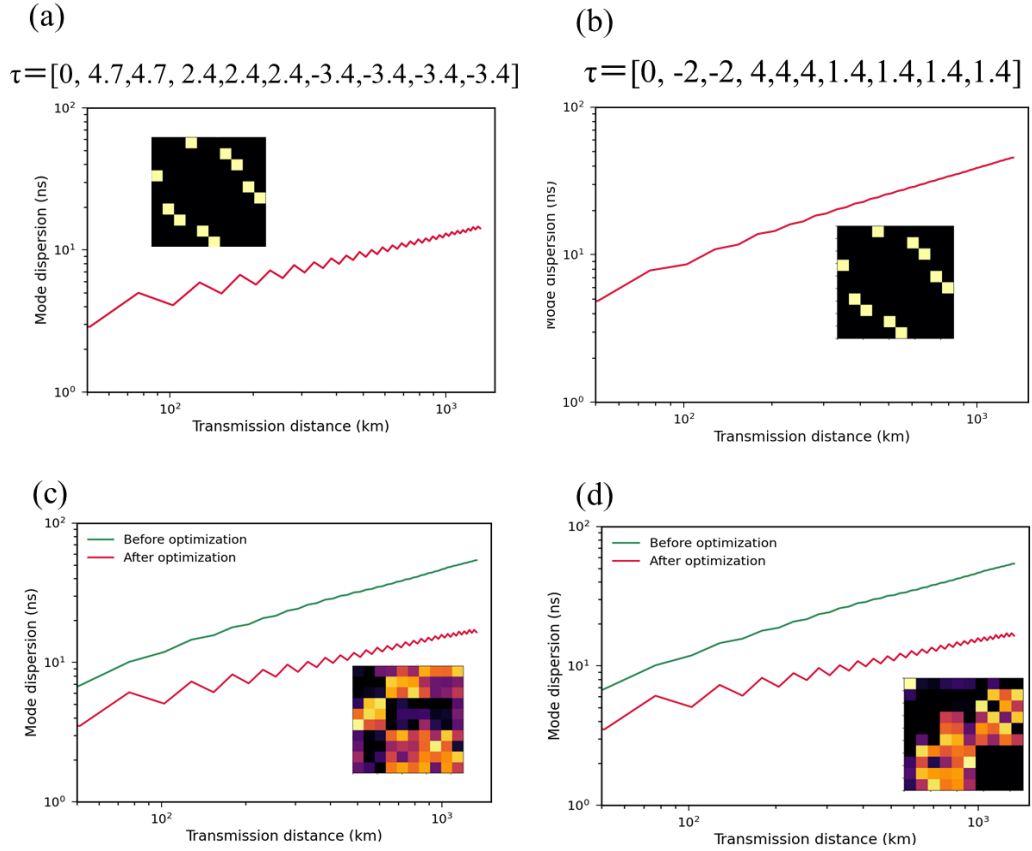

*Figure S9. Dependence of optimal unitary transformation on mode delay in MMF.*

**S10. Comparison with mode scrambling and permutation with passive device**

Previous studies in SDM transmission have reported that passive random mode scrambling can statistically average out differential mode delays, resulting in a square-root dependence on transmission distance rather than linear growth [12,13]. In contrast, our proposed parametrized SDM transmission identifies optimal unitary transformations at each fiber span, allowing us to counteract modal delay accumulation deterministically and achieve stronger suppression than random scrambling alone, which can be confirmed by comparing the square root line and optimized curve shown in Fig. 2(c)–(j). An alternative class of prior work has explored the use of mode permutations — that is, physically reordering fiber connections of the SDM fan-in and fan-out — to reduce the differential mode delay [37,38]. These approaches correspond to constraining the unitary transformation $U_k$ in our model to a fixed permutation matrix. While permutation-based approaches have shown good performance, their optimization is inherently combinatorial and non-differentiable. As such, many of these studies rely on exhaustive or heuristic search strategies, which are computationally expensive and often fail to identify globally optimal solutions, particularly as the number of modes increases. In addition, the passive permutation matrices lack the adaptability for dynamic network reconfiguration. Our simulations show that even when crosstalk characteristics are held constant, the optimal permutation changes with the modal distribution of the modal delay. This suggests that fixed permutation-based systems cannot adapt to varying link conditions such as route changes or the replacement of fiber types. In contrast, our programmable photonic unitary processor enables adaptive reconfiguration, offering the flexibility needed for practical transmission systems.

**S11. Dependence of optimal unitary transformation on mode delay in MMF**
To assess the adaptability of optimized unitary transformations across different mode delay profiles, we evaluated whether a unitary matrix optimized for a given delay condition remains effective under different delay settings. Figure S9(a) and (b) directly illustrate this comparison. In Fig. S9 (a), the mode delay vector is given by $\tau$=[0, 4.7, 4.7, 2.4, 2.4, 2.4, −3.4, −3.4, −3.4, −3.4]/span, which determined the experimental ten-mode fiber delays [S2]. The span length wwas set to 52 km. This matrix performs effectively under the original delay condition, as seen from the suppressed DMD trend. However, Figure S9(b) applies the same optimized permutation matrix originally obtained from the delay profile in (a) to a different delay condition, $\tau$=[0, −2, −2, 4, 4, 4, 1.4, 1.4, 1.4, 1.4]/span. In this case, the same matrix fails to suppress mode dispersion effectively, and DMD accumulates rapidly with distance. This contrast demonstrates that even within the class of structured or group-delayed systems, the optimal permutation depends sensitively on the specific configuration of modal delays. The result underscores a key limitation of static permutation-based approaches: while they can be effective when matched exactly to the system's modal characteristics, they lack the generality required to adapt across different transmission scenarios. Therefore, a fixed-mode mapping strategy—even if optimized once—is not guaranteed to perform well in the presence of fiber-type changes, environmental perturbations, or network reconfigurations.

In contrast, the proposed parameterized SDM framework, which allows continuous control over arbitrary unitary transformations, remains applicable across a broad range of delay conditions [see Figs. S9(c) and (d)]. In these two cases, we applied our optimization framework independently to two distinct mode delay profiles—each substantially different in structure—and obtained optimized unitary matrices tailored to their respective conditions. For both cases, the proposed method significantly reduces the accumulated mode dispersion over long distances. This result highlights a crucial strength of our programmable photonic processor architecture: unlike fixed permutation schemes, the processor can dynamically reconfigure its internal transformation matrix in response to different delay profiles. Whether the delays are symmetric, clustered, or irregularly distributed—as in Fig. S9(c) and S9(d)—the system can always find and apply a near-optimal unitary transformation to suppress DMD.

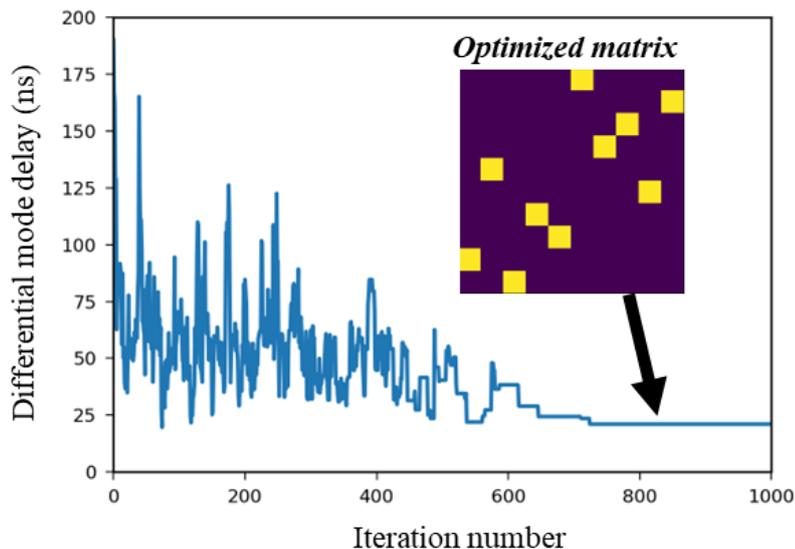

*Figure S10. Optimization using simulated annealing.* *Optimization curve for differential mode delay by using simulated annealing. Inset shows optimized unitary matrix.*

**S12. Optimization method dependency**

To compare the characteristics of different optimization strategies for photonic unitary transformation, we evaluated a gradient-based backpropagation (BP) method and a gradient-free alternative using simulated annealing (SA). Both approaches target the minimization of differential mode delay (DMD), but their performance and computational efficiency differ significantly. Figure S10 shows the DMD convergence behavior during optimization using SA. Due to its heuristic and stochastic nature, the convergence is considerably slower and more erratic than BP. In this ten-mode simulation, the optimization required around 800 iterations to reach a quasi-stable region. The minimum DMD achieved using SA was slightly worse than that obtained via BP-based optimization. In contrast, the BP method—discussed in the main text—exploits the differentiability of the transmission model to perform gradient-based updates on continuous unitary parameters. As a result, it converges ~1/10 the number of iterations and consistently reaches lower DMD values. This is attributed in part to the ability of BP to explore the full continuous space of unitary matrices, rather than the restricted discrete configurations (e.g., permutation matrices) typically accessible to SA. That said, the underlying reason for the performance gap remains an open question. It is not yet clear whether the observed difference arises from the intrinsic limitation in the representational flexibility of discrete unitary forms (as used in SA) or from the tendency of the SA algorithm to fall into local minima in a rugged optimization landscape. Addressing this question will require further investigation.

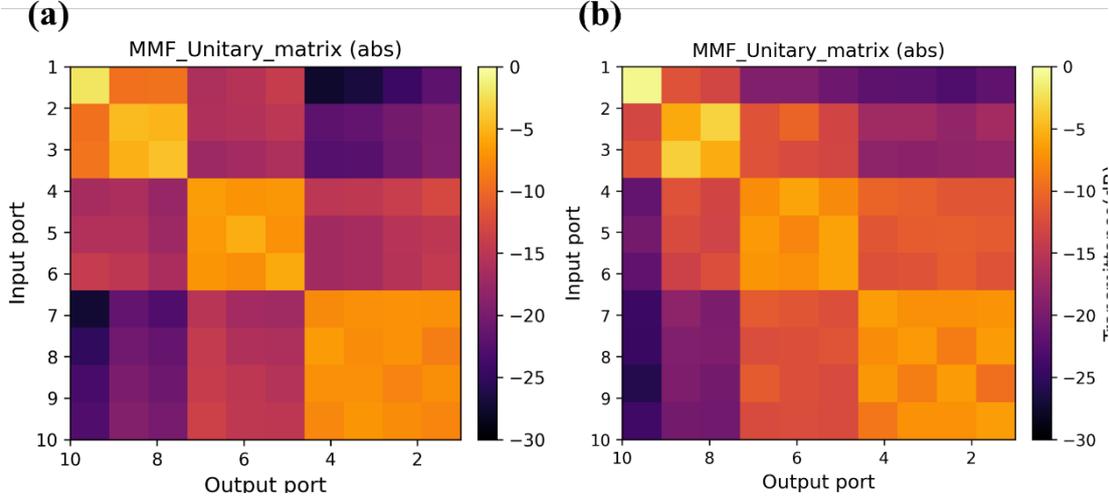

*Figure S11. Generation of unitary matrix emulating FMM.* (a) Generated and (b) experimental [S2] unitary matrix for ten-mode fiber using proposed scheme.

### S13. Generation of arbitrary shape unitary matrix

In order to simulate physically realistic mode coupling behavior in multimode fiber (MMF) transmission systems, we developed a structured method for generating unitary matrices that reflect group-wise modal interactions. This approach allows us to emulate both strong intra-group coupling and weak inter-group crosstalk while maintaining the overall unitarity constraint. The procedure is outlined below:

1. **Initialize block-wise strong coupling structure:**

    Each mode group (e.g., corresponding to a degenerate LP mode set) is assigned a block of strong internal coupling. Within each group, a random unitary matrix is generated using Haar sampling to reflect fully mixed modal interactions. These blocks are combined along the diagonal of a complex matrix $A$, forming an initial "intra-group coupling" structure.

2. **Introduce inter-group weak coupling:**

    Weak crosstalk between adjacent mode groups is introduced by assigning small random complex values to the off-diagonal sub-blocks of matrix $A$. The magnitude and sparsity of these off-diagonal entries control the strength and range of modal interactions, enabling fine-grained modeling of weakly coupled systems. This process may be repeated for second-nearest groups and beyond to simulate extended coupling profiles.

3. **Orthogonalization via QR decomposition:**

    The resulting matrix $A$, which encodes both strong intra-group and weak inter-group modal coupling, is then orthogonalized via QR decomposition: $A=QR$, where $Q$ is a unitary matrix and $R$ is upper triangular. To ensure that the unitary matrix $Q$ retains the relative magnitude profile (i.e., intensity coupling characteristics) encoded in $A$, we preserve the absolute values of $Q$ as inherited from $A$.

This method produces unitary matrices that preserve realistic spatial mode dynamics—strong within degenerate groups and weak but structured across group boundaries—while maintaining mathematical rigor and computational efficiency. The statistical distributions of the resulting unitary matrices are shown in Fig. S11. This physically informed modeling approach provides a practical foundation for constructing the transmission matrices $S_k$ and $V_k$ used in our simulations of SDM systems. It enables a broad exploration of fiber conditions while retaining relevance to experimentally observed mode coupling behavior.

[S1] R. Ospina et al., OFC2020, W2A.47, 2020

[S2] 37.  Shibahara, K., Hoshi, M., and Miyamoto, Y. "10-spatial-mode 1300-km transmission over 6-LP graded index few-mode fiber with 36-ns modal dispersion." *J. Light. Technol.* 42(4), 1257-1264 (2024).